# Human-like working memory signatures emerge from intrinsically plastic artificial neurons for robust dynamic vision

Jingli LIU[1], Huannan ZHENG[1], Bohao ZOU[1], Kezhou YANG[1]*

[1]Microelectronics Thrust, The Hong Kong University of Science and Technology (Guangzhou), Guangzhou, 511453, China

*E-mail: kezhouyang@hkust-gz.edu.cn

## Abstract

Processing dynamic real-world sensory streams requires mechanisms that retain task-relevant recent context while limiting energy consumption and the accumulation of obsolete information. Conventional recurrent and spatiotemporal architectures encode temporal context through recurrent hidden states, temporal kernels, or explicit input windows. Biological working memory, in contrast, is thought to rely on interacting activity-dependent states spanning neural activity, synaptic efficacy, and intrinsic cellular dynamics. Inspired by activity-silent accounts of working memory, we introduce Intrinsic Plasticity Network (IPNet), a neuromorphic architecture in which temporal context is encoded by a neuron-level intrinsic state, without recurrent connectivity, synaptic memory variables, or explicit network-level history buffers. We physically instantiate this mechanism using Joule-heating-induced dynamics in magnetic tunnel junction (MTJ) neurons, whose volatile internal states retain a transient trace of recent input history. The resulting memory exhibits signatures associated with biological working memory, including limited effective capacity, interference effects and a decaying recency effect. Using physical parameters directly extracted from experimental MTJ devices, IPNet outperformed the evaluated full-precision digital temporal architectures spanning recurrent, spatiotemporal-convolutional and attention-based models, including LSTM, R(2+1)D and Video Swin baselines, on temporally sensitive DVS recognition tasks, and reduces steering prediction errors by 12.4% in continuous autonomous driving scenarios under comparable backbone settings. We further executed the temporal-memory layer using fabricated MTJs in a hardware-in-the-loop system, where physical-device outputs closely reproduced the calibrated model and retained an advantage over a state-dimension-matched full-precision software LSTM. By embedding temporal state directly in device physics, IPNet is projected to require a neuron footprint of approximately 1.5 $\mu m^2$, while device-level estimates indicate substantially lower memory-related temporal-processing energy than the evaluated GPU-based baselines (up to 90,000×). These results indicate that physics-grounded, biologically inspired latent-state dynamics can enable robust and energy-efficient temporal processing.

## Main

Dynamic perception requires integrating recent task-relevant information while limiting interference from obsolete inputs. Biological working memory offers a functional principle for this balance: it selectively maintains and updates a limited amount of information over short periods, rather than preserving a lossless sensory history[1–3]. At the neural level, evidence and models implicate multiple, potentially complementary mechanisms, including persistent and intermittent population activity, short-term changes in synaptic efficacy and activity-dependent changes in intrinsic neuronal excitability[4–9]. Such finite and temporary retention may be particularly advantageous in continuous sensory environments, where tasks such as motion estimation and vehicle steering depend on recent temporal context[10,11]. Modern artificial intelligence systems, in contrast, typically maintain temporal context through recurrent connections[12], parallel spatiotemporal inputs[13] or attention mechanisms[14,15] that buffer extensive digital histories. Although effective, these mechanisms require recurrent state updates, multi-frame computation or the storage and movement of sequence representations; their resource costs grow with the temporal context considered[16], and can become vulnerable to the accumulation of historical noise in continuous real-world sensory streams[17–19].

Activity-silent accounts of working memory provide a biological principle for addressing this challenge. Rather than requiring information to be continuously represented by persistent neural firing, these accounts propose that recent information can be retained in transient, activity-dependent latent states[7,20–22]. In computational neuroscience, biophysically grounded models commonly attribute such latent states to short-term changes in synaptic efficacy[8,23]. In hardware implementations that assign a dynamic state variable to each synapse, however, the number of state elements and updates scales with network connectivity. Neuronal intrinsic plasticity (Fig. 1a), a mechanism enabling short-term information retention independently of synaptic connections[4,5], suggests a simpler neuron-level mechanism. Activity-dependent changes in neuronal excitability can persist after the eliciting input and influence responses to subsequent stimuli[9,24]. This neuron-local state can therefore carry recent input history across multiple subsequent responses, without requiring a separate short-term state at each synapse.

Translating this memory principle into an efficient physical implementation remains challenging. To achieve brain-level power efficiency, the neuromorphic computing paradigm proposes emulating these biological mechanisms. Spiking neural networks[25] (SNNs), particularly when processing asynchronous events from dynamic vision sensors[26,27] (DVS), offer a theoretically energy-efficient framework inspired by biological systems. However, when implementing short-term memory at the neural network level, to ensure competitive performance, current SNNs frequently follow complex non-spiking ANN architectures like LSTMs[28] or Transformers[29], which favors the compute-dense nature of GPU-based systems rather than the efficiency of the biological brain[29–31]. Furthermore, emulating complex temporal dynamics in standard silicon incurs a critical hardware penalty. For instance, upgrading a standard Leaky Integrate-and-Fire (LIF) neuron to an adaptive variant (ALIF) necessitates auxiliary feedback loops, effectively may double the circuit footprint and offsetting the scalability gains of neuromorphic systems[32,33]. Thus, although neuronal intrinsic plasticity provides a compact biological principle, emulating it through explicit circuits can offset the scalability advantages of neuromorphic systems. Emerging nanodevices offer a different route by allowing the required temporal state to arise directly from their intrinsic physical dynamics[34].

By combining this bio-inspired neuron-level memory mechanism with the intrinsic thermal dynamics of magnetic tunnel junctions, we introduce IPNet, a neuromorphic architecture in which all temporal memory is carried by neuronal intrinsic plasticity. At the network level, IPNet requires no recurrent connectivity, synapse short-term memory states, or explicit history buffers. We implement the neuronal core using Magnetic Tunnel Junctions (MTJs), repurposing their intrinsic Joule-heating dynamics, which are conventionally suppressed as parasitic noise[35,36], into the physical source of plasticity. Recent inputs alter the internal thermal state of an MTJ neuron, which transiently modulates its responses to subsequent inputs before gradually relaxing toward equilibrium. This device-local state therefore encodes recent input history and produces a finite and decaying memory trace without per-synapse memory devices or additional memory-control circuitry. Furthermore, we demonstrate that this physically derived working memory confers a direct functional advantage in dynamic visual processing because of its biological-like constraints. Its limited capacity and transient decay act as a temporal filter that retains recent task-relevant context while suppressing redundant or noisy historical information. By offloading temporal processing directly to MTJ device physics, IPNet eliminates the need for recurrent connections, synapse-specific memory variables, explicit history buffers and auxiliary memory-control circuits (Fig. 1a). This device-intrinsic memory mechanism enables a compact estimated neuron footprint of approximately 1.5 $\mu m^2$. Across the evaluated configurations, IPNet reduces the estimated memory-related temporal-processing energy by 100–90,000× and the classification error by up to 35× relative to the evaluated GPU-based temporal baselines on the Time-Reversed DVS Gesture task (Fig. 1b). Using parameters directly extracted from experimental MTJ devices and under comparable backbone settings, IPNet outperforms conventional temporal-processing architectures, including LSTM-based models, R(2+1)D[13] and a pretrained Video Swin Transformer[15], on temporally sensitive DVS recognition tasks. In continuous event-based driving scenarios, IPNet further reduces steering-prediction errors by 12.4% compared with a matched recurrent baseline. This work establishes a physics-driven neuromorphic paradigm in which the biological constraints of working memory are transformed into a scalable computing principle for robust and energy-efficient dynamic perception.

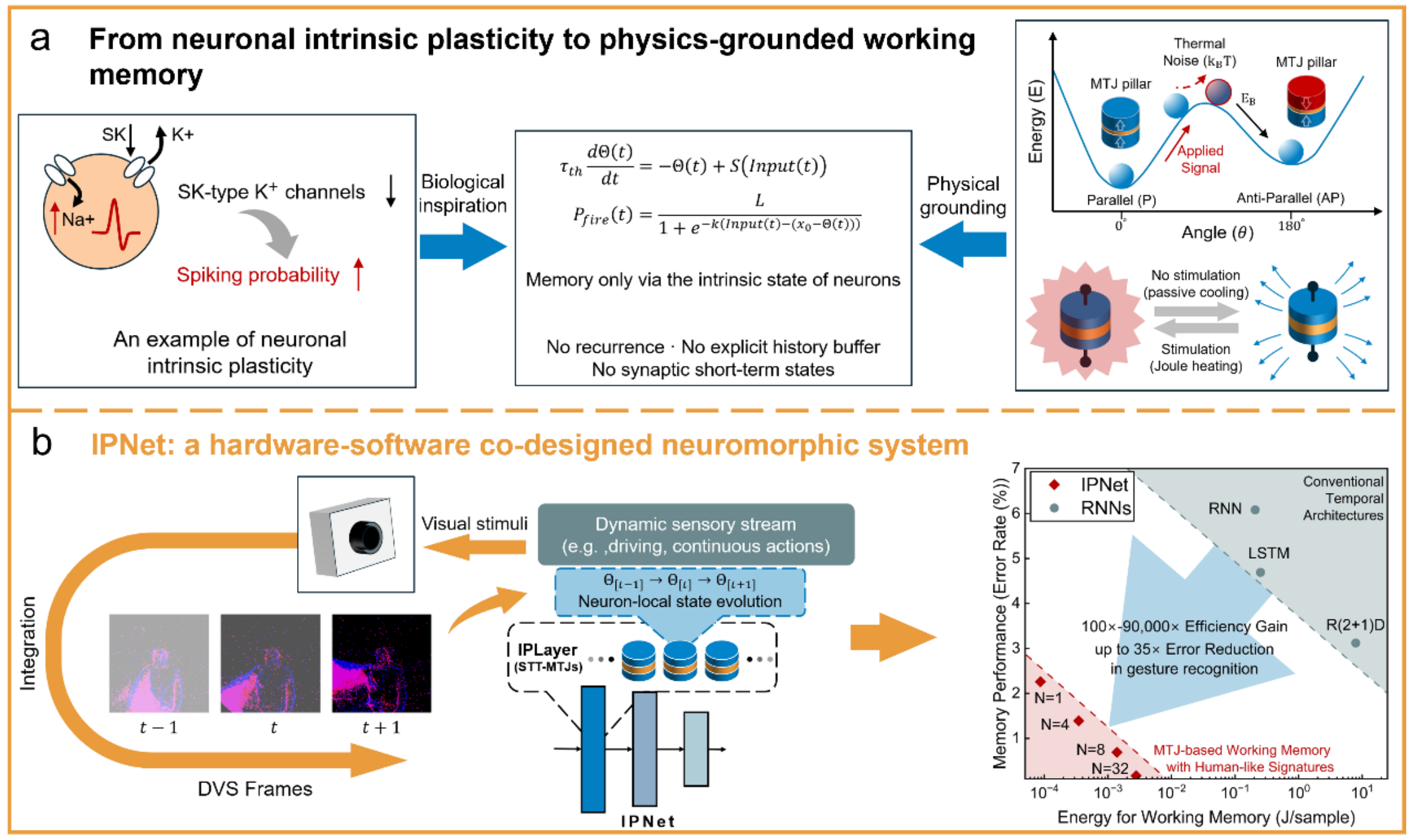

**Fig. 1. Physics-grounded neuronal intrinsic plasticity enables efficient working memory in IPNet. a,** Neuronal intrinsic plasticity motivates a neuron-local state, $\Theta(t)$, that retains recent inputs and modulates firing probability. In MTJ-based neurons with intrinsic plasticity, this state is physically implemented through Joule-heating-induced thermal dynamics, with stimulation enhancing thermal fluctuations and passive cooling

allowing the state to decay. This mechanism requires no recurrent connectivity, explicit history buffer or synaptic short-term states. **b,** Hardware-software co-designed IPNet for dynamic sensory processing. Sequential DVS frames are processed by an STT-MTJ-based IP layer with evolving neuron-local states. Right, classification error versus estimated memory-related energy per sample on the Time-Reversed DVS Gesture task. $N$ denotes the number of thermally sensitive MTJs per computing unit. IPNet achieves up to a 35-fold reduction in classification error and a 100- to 90,000-fold energy-efficiency gain over the evaluated GPU-based baselines.

## Thermally driven intrinsic plasticity in MTJ-based stochastic neurons

To physically realize the transient dynamics of biological working memory, we repurposed the inherent thermodynamic properties of standard magnetic tunnel junctions (MTJs) (Fig. 2a, c). By interfacing the MTJ with a comparator circuit (Fig. 2d), its probabilistic magnetic moment switching due to spin-transfer torque (STT)[37,38] is translated into the sigmoidal activation profile of a stochastic neuron (Fig. 2e). Crucially, during the switching process of STT-MTJ operation, the applied current passes through the device. As a result, Joule heat is an inevitable byproduct. In our proposed neuron model, we harness Joule heat as a mechanism for neuronal intrinsic plasticity. Upon stimulation, device temperature is increased by Joule heat, which enhances the thermal fluctuations. These enhanced thermal fluctuations directly modulate the switching probability of the MTJ[39,40]. At the neuron functional level, this manifests as a leftward shift in the activation curve, effectively reducing the firing threshold and increasing the firing probability for a certain input. This transient shift in the device probabilistic firing curve, mimicking neuronal intrinsic plasticity, can serve as an information storage mechanism[4,5] for working memory in neural networks.

To quantify this effect, we emulated the thermal residue of preceding input by introducing a pre-heating pulse of varying duration prior to the standard measurement pulse. It was found that the inherent Joule heating generated during standard STT operations is sufficient to induce profound modulations in firing probability. For instance, the firing probability nearly triples (from 9.5% to 27.4%) immediately following a 500 ns, 0.9 V pre-pulse, before passively relaxing to 16.3% after a 200 ns cooling interval. Meanwhile, to construct a hardware-calibrated neuron model (see Methods), we parameterized the neuronal state using the sigmoid mean ($x_0$, the voltage amplitude corresponding to half of the maximum switching probability). Experimental data confirms that the MTJ's probabilistic activation across dynamic thermal states can be captured by the change in the shift of sigmoid function mean $x_0$ (Fig. 2f, g).

We further validated this *in-situ* temporal processing under continuous spiking conditions (230 ns cycle). Modeled as a dynamic variable, the neuron's sensitization state ($x_0$) exhibits approximately exponential decay with the pulse interval and is primarily governed by the amplitude, interval and sequence of preceding events (Figs. S1-S3). To validate the fidelity of our modeling approach, we benchmarked the software simulations against experimental data, observing a high degree of quantitative agreement (Fig. S4). Such fidelity confirms that temporal processing occurs in-situ and concurrently with spiking activity, removing the need for an explicit temporal-state update phase in the proposed neuron model. Ultimately, this design successfully repurposes parasitic Joule heating into a fundamental computational resource, physically instantiating intrinsic plasticity without auxiliary memory components.

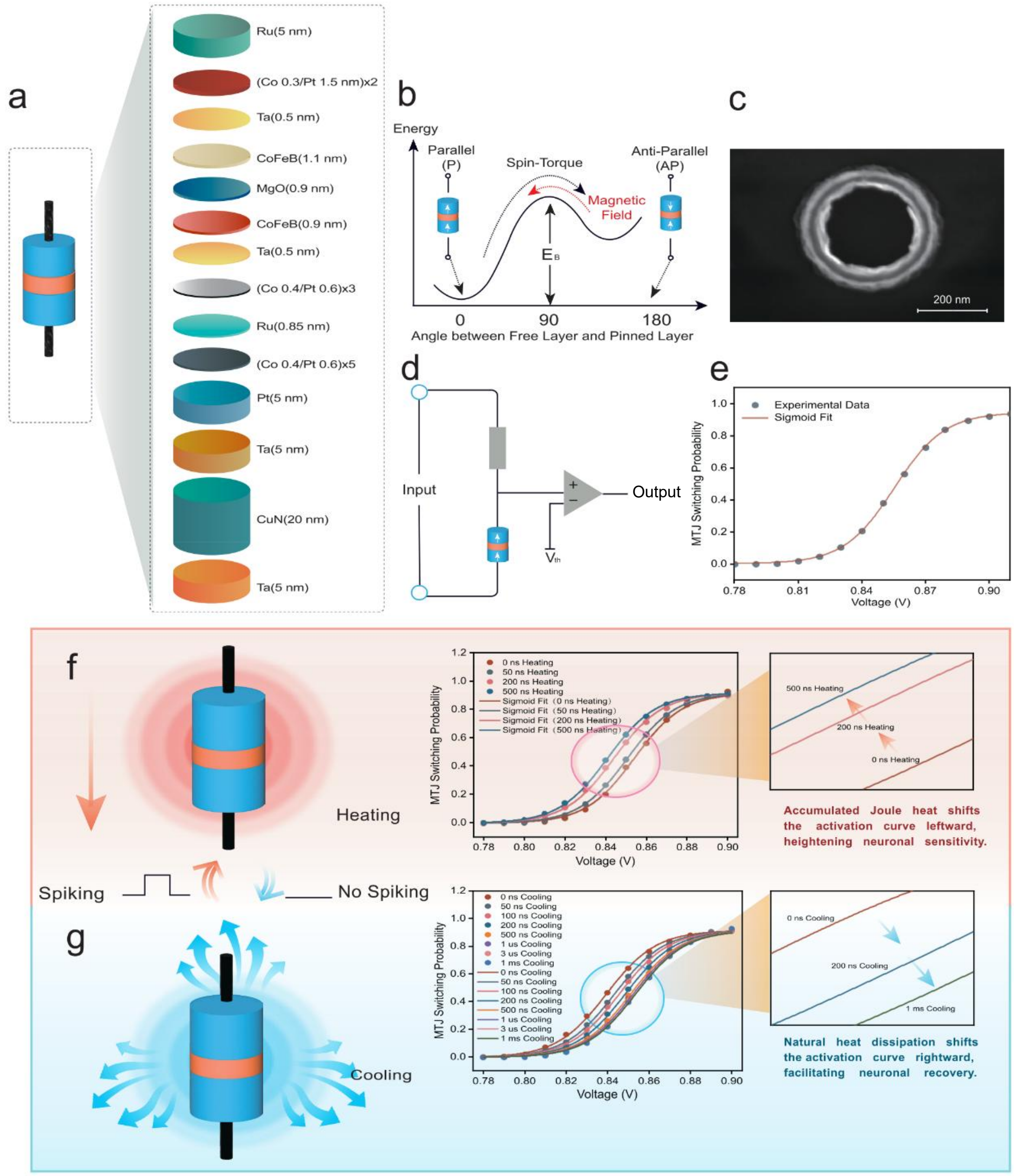

**Fig. 2. Device structure and characterization of the MTJ-based stochastic intrinsic plasticity neuron. a**, Schematic illustration and detailed multilayer stack composition of the MTJ used in this work. **b**, Energy landscape illustrating the switching mechanism between parallel (P) and anti-parallel (AP) states driven by spin-transfer torque (STT).

**c**, Top-view SEM image of the fabricated MTJ nanopillar with a diameter of approximately 200 nm. **d**, Simplified circuit schematic of the MTJ-based stochastic neuron. **e**, Switching probability of the MTJ neuron as a function of input voltage, showing experimental data (gray filled circles) and the sigmoid fit (red solid line). **f**, Sensitization of the MTJ neuron induced by Joule heating. The activation curves exhibit a systematic leftward shift as the pre-heating pulse duration increases. **g**, Cooling and recovery dynamics of the MTJ neuron. The activation curves gradually shift back to the right over time in the absence of input, indicating a return to the initial state. In both **f** and **g**, solid circles represent experimental data, while solid lines denote sigmoid fits obtained by varying only the sigmoid mean ($x_0$) while keeping other parameters constant.

## Emergent human-like working memory features from neuronal intrinsic plasticity

To investigate whether single-neuron thermal plasticity can generate network-level memory dynamics, we developed the Intrinsic Plasticity Network (IPNet). In a minimal two-layer fully-connected architecture (Fig. 3d), which was chosen to isolate the contribution of intrinsic neuronal states from architectural memory mechanisms, temporal memory emerges exclusively from neuronal thermal states rather than recurrent synapses or explicit history buffers. Using a differential readout (Figs. S8, S9) that pairs thermally plastic neurons with insensitive reference units to better capture changes in the neurons' states, we benchmarked IPNet against a standard long short-term memory (LSTM) network (see Methods).

Across three classic cognitive tasks, IPNet exhibited several constraints observed in human working memory behavior, whereas the LSTM, as expected for an architecture with an explicit gated memory state, maintained near-perfect retention. In the *n*-back task[41] (Fig. 3a), IPNet performance progressively declined as memory load (*n*) increased (Fig. 3b), following the load-dependent trend observed in human behavioral data[42,43]. This capacity limit arises naturally from device heat dissipation; in contrast, the same feed-forward architecture with standard LIF neurons, but without intrinsic plasticity or recurrent connections, failed the task.

IPNet also reproduced human-like error patterns: proactive and retroactive interference[44–46] increased with target-interferer similarity (Fig. 3c,e), using the distributed and orthogonal stimulus representations shown in Supplementary Fig. 16. A retrieval cue mitigated retroactive interference, resembling cue-dependent recovery in human working memory[47] (Fig. S5). Finally, in a classic serial-position free-recall task[48] (Fig. S17), IPNet displayed a decaying recency effect with little primacy effect(Fig. 3f, g), consistent with a transient short-term memory trace rather than long-term memory[48].

The divergence between LSTM and IPNet highlights a key computational distinction: IPNet retains recent history through a finite, decaying and interference-sensitive physical state rather than a high-fidelity digital buffer. These working-memory-like constraints may be advantageous in dynamic sensory environments, where preserving all past inputs can accumulate obsolete or noisy context. We therefore next test whether the same intrinsic physical memory can act as a temporal filter for robust dynamic perception.

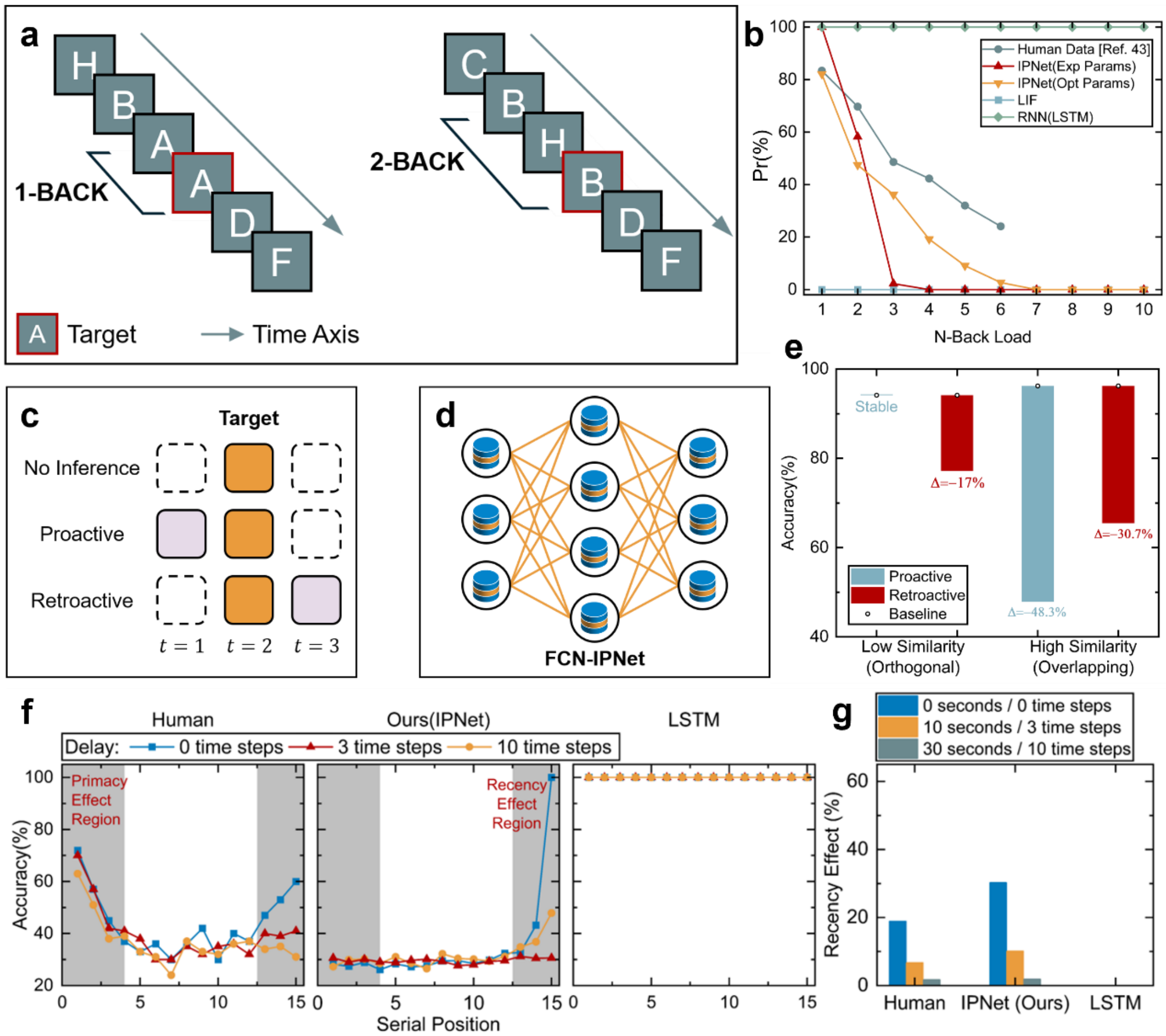

**Fig. 3. Human-like working memory constraints emerging from neuronal intrinsic plasticity. a**, Schematic illustration of the n-back task, using 1-back and 2-back conditions as examples. Participants are required to determine whether the current stimulus matches the one presented $n$ steps earlier. **b**, N-back performance as a function of memory load for IPNet using experimental (Exp.) or optimized (Opt.)

device parameters, a feedforward LIF network, an LSTM and human participants[43]. **c**, Experimental protocol for memory interference. Interference paradigm. The target is presented at $t = 2$, with a distractor either preceding it (proactive interference, $t = 1$), following it (retroactive interference, $t = 3$), or absent (no interference); dashed boxes indicate no input. **d**, Schematic of the two-layer fully connected IPNet used for the working-memory tasks. **e**, Relative accuracy degradation of the IPNet under proactive and retroactive interference compared to the control group, tested with low (orthogonal) and high (overlapping) similarity between target and distractor. **f,** Serial-position curves for humans[48], IPNet and LSTM after retention intervals of 0, 3 and 10 model time steps, corresponding to 0, 10 and 30 s in the human data. Shaded regions indicate the primacy and recency regions. The LSTM curves overlap at 100% accuracy. **g**, Decay of the recency effect over time. The recency effect (%) is calculated as the difference between the average accuracy of the last 3 serial positions (recency region) and the middle 5 positions (to minimize the influence of primacy and recency effects).

## Constrained MTJ-based physical memory improves dynamic vision

Unlike explicit digital memory architectures such as LSTMs, IPNet exhibited several human-like working memory behavioral across controlled memory tasks. We next asked whether these biologically inspired constraints—limited capacity, decaying retention and interference sensitivity—could provide a functional advantage in dynamic visual processing by filtering obsolete or redundant historical context rather than preserving all past inputs, similar to human behavior[49,50]. To test this, we constructed IPNet18, an all MTJ-neuron based spiking neural network on an adapted

ResNet18 backbone[51,52] (see Methods). This network has simple 2D feed-forward architecture (Fig. 4d), eliminating the need for recurrent synapses and multi-frame parallel inputs commonly used for temporal processing in dynamic vision. Instead, all history-dependent temporal processing in IPNet is carried by the intrinsic physical memory of thermally plastic MTJ neurons. Thus, IPNet tests whether a finite, decaying physical memory can replace explicit recurrent, convolutional or attention-based temporal aggregation in dynamic vision.

We first used the standard 11-class DVS Gesture dataset[53] as an initial benchmark before constructing a temporally controlled variant. IPNet18 achieved 99.65% accuracy, exceeding the spatial ResNet18 baseline and the evaluated temporal baselines, including LSTM-, R(2+1)D[13]- and Video-Swin[15]-based models (Fig. 4c). Replacing the intrinsic-plasticity neurons with standard LIF neurons reduced accuracy to 97.57%, indicating that the gain depends on the MTJ-derived intrinsic state. This performance was obtained using experimentally measured MTJ parameters and without software-level data augmentation. However, the strong performance of the purely spatial ResNet18 baseline suggests that standard DVS Gesture is dominated by spatial cues and does not isolate the requirement for temporal processing.

To emphasize the temporal processing capabilities, we introduced a more demanding 'Time-Reversed DVS Gesture task', where the chronological order of the event stream is strictly inverted while spatial coordinates are preserved (Fig. 4a). Because each original gesture and its reversed counterpart (22 classes in total) share the same accumulated spatial events, successful classification requires sensitivity to temporal information rather than static event position. The spatial-only ResNet18 baseline dropped from 97.22% on the original DVS Gesture task to 48.61% on this temporally controlled task (Fig. 4c). This halving of accuracy indicates that the spatial model can still recognize the underlying gesture morphology, but lacks access to temporal - dependent cues and therefore tends to confuse each gesture with its time-reversed counterpart.

In this temporally sensitive task, IPNet showed minimal accuracy degradation, whereas the evaluated temporal baselines exhibited more pronounced performance drops. The accuracies of recurrent architectures, including RNN and LSTM, and the spatiotemporal convolutional model R(2+1)D decreased from 97.22-97.92% on the original DVS Gesture task to 93.92-96.88% on the Time-Reversed DVS Gesture task (Fig. 4c). In contrast, IPNet18 maintained a robust 99.31% (originally 99.65%) accuracy, achieving the highest accuracy among the evaluated baselines. We further included a pretrained Video Swin Transformer as an attention-based spatiotemporal baseline. Although its absolute accuracy is likely affected by the limited scale of DVS Gesture and the domain mismatch between natural-video pretraining and event-based inputs, it decreased from 97.22% to 96.35% after temporal reversal, a larger degradation than IPNet. These results suggest that IPNet's advantage does not arise from preserving a longer history, but from using a finite physical memory trace that emphasizes recent event transitions while allowing obsolete activity to decay.

Visualization of the internal representations further explains this temporal-discrimination capability (Fig. 4b). In IPNet18, the intrinsic plasticity of MTJ neurons converts differences in event order into separable temporal-dependent feature representations, producing a clear structural divergence between the original and time-reversed streams. When intrinsic plasticity was ablated, original and reversed streams produced nearly identical feature patterns, resulting in identical spatial patterns that fail to encode the temporal inversion, yielding an accuracy of ~50%.

We next tested whether these working-memory-like constraints support temporal generalization when only limited input history is available. When trained with the first 20% frames in each sample of the original dataset, RNN, LSTM, and R(2+1)D can achieve accuracies of 96.18%, 96.18%, and 96.88%, respectively (Fig. S13). However, performance on the reversed dataset reveals critical limitations. RNN and LSTM suffer significant drops to 68.92% and 78.30%, indicating that the reduction in frame count severely impacts their temporal processing. R(2+1)D maintained a high accuracy of 96.53% under the sliding-window protocol, but this robustness depended on the windowed aggregation strategy; when the sliding window was removed, its accuracy decreased to 89.41% on the reversed partial dataset. In contrast, IPNet18 maintained an accuracy of 99.48%, comparable to its full-sequence performance of 99.31%, indicating more robust temporal generalization under shortened event streams. Moreover, IPNet performance scales with the number of thermal sensitive MTJs per computing unit ($N$) (preceding result used $N = 16$) (see Methods). Further increasing $N$ to 32 raises the accuracy to a maximum of 99.83% (Fig. S11).

Furthermore, the physical memory exhibited a pronounced "Memory-at-the-Frontier" effect, in which temporal processing was most effective when the intrinsic-memory module was placed close to the sensory input. Accuracy was maximized when the MTJ-based memory module was inserted at the input layer, reaching 99.31%. Moving memory module after the first convolutional layer reduced the accuracy to 95.66%, whereas placing it immediately before the output layer caused the performance to fall to approximately 50% (Fig. 4e). This strong placement dependence indicates that intrinsic physical memory acts most effectively as a near-sensor temporal filter, where fine-grained event-order information is still available. Once the event stream has been progressively abstracted by deeper spatial feature extraction, the original temporal-dependent cues are largely compressed or lost, limiting the benefit of late-stage memory insertion.

To directly examine how intrinsic-state decay shapes temporal retention and temporal-order discrimination, we varied the decay scale of simulated MTJ neurons. In the delay-matching task, smaller decay scales maintained high accuracy over longer delays, showing that the decay dynamics determine the persistence of the retained temporal trace (Fig. S14a). By contrast, on the Time-Reversed DVS Gesture task, models trained using the first 20% of each sequence and evaluated on the complete sequence showed monotonically higher accuracy with increasing decay scale, from 88.89% at a decay scale of 0 to 93.06% at 0.5 (Fig. S14b). Thus, prolonging the intrinsic state did not improve temporal discrimination under this protocol. Persistent activity from earlier events may instead interfere with the representation of later transitions. These results show that the advantage of IPNet does not arise from simply maximizing physical memory duration. Instead, it depends on a finite and task-matched memory trace that preserves recent event transitions while allowing outdated activity to dissipate.

Importantly, the layer composed of neurons with intrinsic plasticity, hereafter referred to as the IP layer, can also be integrated into non-spiking artificial neural networks. Integrating the IP layer into a conventional non-spiking ConvNeXt v2-tiny architecture[54] improved its performance on two challenging event-based recognition datasets. Specifically, this integration improved accuracy on the Dailydvs-200 dataset[55] from 45.08% to 47.88%, and on the HARDVS dataset[56] from 52.18% to 53.19% (see Methods). We further coupled the IP layer to a Swin Transformer-Tiny backbone without explicit spatiotemporal attention. On the Time-Reversed DVS Gesture task, this architecture achieved an accuracy of 97.22%, exceeding the 96.35% obtained by the Video Swin Transformer-Tiny baseline. Thus, a compact intrinsic state introduced at the neuron level can provide temporal sensitivity to otherwise spatial CNN and Transformer backbones, extending the applicability of the proposed mechanism beyond spiking architectures.

Crucially, IPNet achieves this temporal-order processing with minimal additional parameters and substantially reduced memory-related temporal-processing energy. The estimated energy consumption of all IP neurons in IPNet18 is approximately 87 μJ per DVS Gesture sample, including both stochastic pulse sampling and all intrinsic-plasticity-based temporal processing. This estimate was calculated using the minimal configuration of $N = 1$ thermally sensitive MTJ neuron per computing unit. Although this configuration does not correspond to the best-performing IPNet setting, it still achieved 97.74% accuracy on the Time-Reversed DVS Gesture task (Fig. S11), exceeding other evaluated baseline models. Processing a single DVS Gesture sample requires only ~87 μJ. When isolating the energy cost specifically attributed to temporal processing, under comparable ResNet18-based backbone, IPNet18 is approximately 2,414×, 2,874×, and 90,920× more energy-efficient than the temporal components of RNN, LSTM, and R(2+1)D models, respectively (see Methods). Together, these results indicate that device-intrinsic, finite memory provides a more efficient and robust substrate for dynamic vision compared with explicit temporal-memory or spatiotemporal aggregation mechanisms.

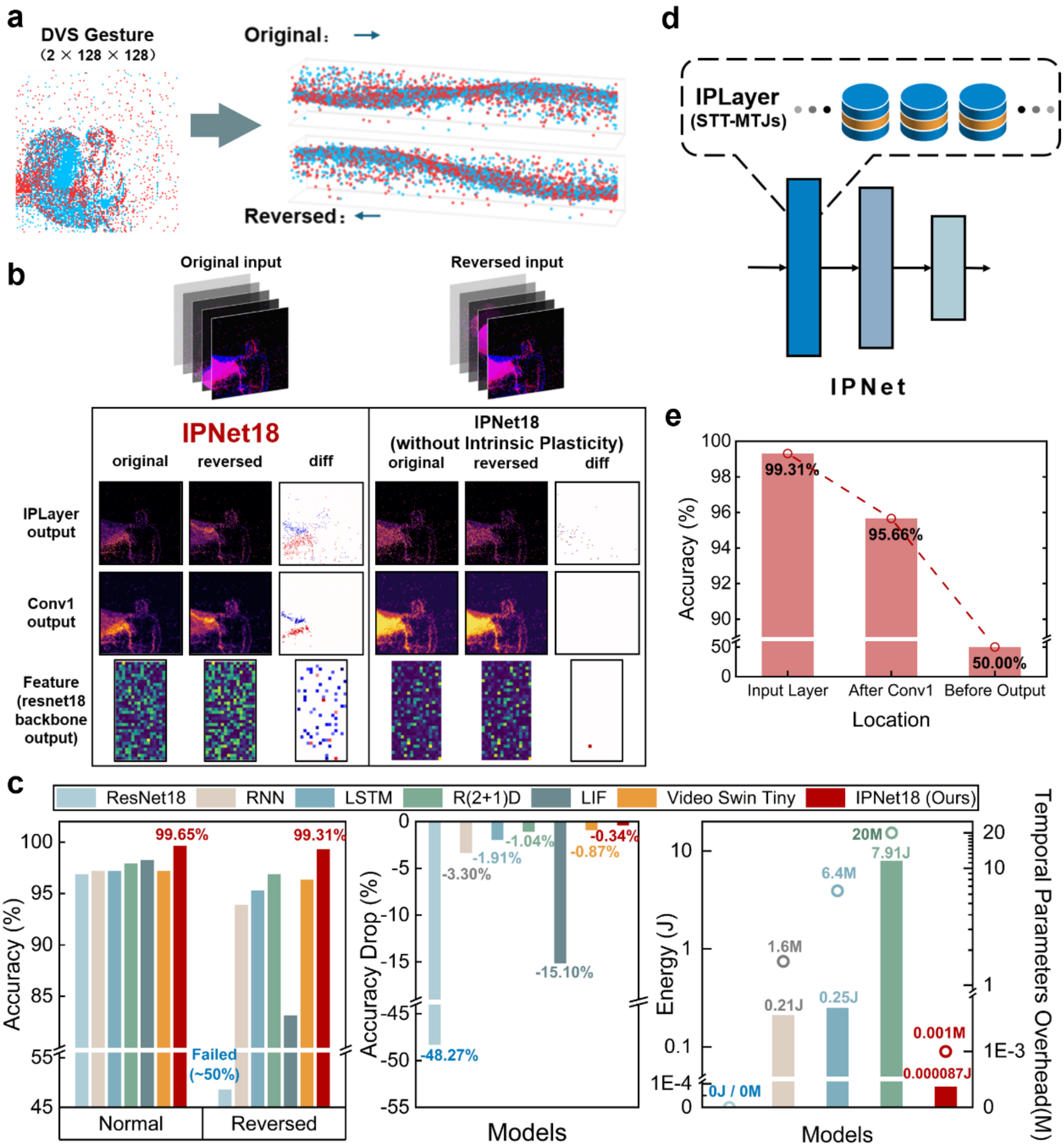

**Fig. 4. Constrained MTJ-based intrinsic memory enables robust and efficient temporal processing in dynamic vision. a**, Construction of the time-reversed DVS Gesture task. The temporal order of each event stream is reversed while the spatial coordinates and event polarities are preserved, such that the original and reversed samples contain the same accumulated spatial information but differ in temporal order. **b**, Representative internal representations generated by IPNet18 and by the corresponding network without intrinsic plasticity. Contrast between feature maps of

IPNet18 (left) and the plasticity-ablated baseline (right) given original versus time-reversed inputs. The 'diff' columns highlight that intrinsic plasticity creates a structural divergence necessary for distinguishing temporal direction, whereas the ablated model produces identical features. **c**, Comparison of classification performance and temporal-processing overhead across ResNet18, RNN, LSTM, R(2+1)D, LIF, Video Swin-Tiny and IPNet18. Left, classification accuracy on the original ('Normal') and time-reversed DVS Gesture tasks. Middle, change in accuracy between the original and time-reversed tasks; negative values indicate reduced accuracy after temporal reversal. Right, estimated incremental energy per sample for temporal processing (bars, left axis) and additional temporal parameter count relative to the spatial-only ResNet18 backbone (open circles, right axis). **d**, Schematic of IPNet18, which comprises a single IP layer containing thermally sensitive MTJs, while all remaining layers use thermally insensitive MTJs as their neuronal cores. **e**, Accuracy degradation on the Time-Reversed DVS Gesture task as the position of the memory-endowing IP layer shifts to deeper layers.

## Working-memory-like constraints improve end-to-end steering prediction

Having established that MTJ-based neuronal intrinsic plasticity gives rise to capacity-limited, interference-sensitive and decaying memory traces in controlled cognitive tasks, and that these traces support temporal-order discrimination in event-based recognition, we next asked whether the same working-memory-like constraints remain advantageous in a continuous control setting. End-to-end driving imposes substantially different computational demands from gesture classification: the model must continuously regress steering commands from noisy and non-stationary sensory streams.

We evaluated IPNet on the DAVIS Driving Dataset 2020 (DDD20) [57], an event-based driving dataset containing more than 51 h of recordings over approximately 4,000 km of highway and urban driving. Following the data-generation and evaluation procedures of Ref. 57, we used only the DVS event stream and trained the models to predict either the instantaneous steering angle or the steering angle 0.2 s into the future under daytime and nighttime conditions. Under matched backbone and training settings, IPNet18 was compared with a spatial ResNet18 without temporal memory and a ResNet18-LSTM incorporating a conventional recurrent state[28,58]. Performance was evaluated using root mean square error (RMSE) and explained variance (EVA) [57,59].

Under both daytime and nighttime conditions, and for both instantaneous and 0.2-s-ahead steering prediction, IPNet18 consistently outperformed the two baseline models (Fig. 5b). Averaged over these four combinations of illumination condition and prediction horizon, IPNet18 reduced RMSE by 13.6% relative to the memoryless ResNet18 baseline. By comparison, ResNet18-LSTM reduced RMSE by only 1.4% relative to the same spatial baseline, and its mean RMSE remained 12.4% higher than that of IPNet18. The EVA scores showed the same overall ranking.

The marginal improvement of ResNet18-LSTM over the purely spatial baseline is consistent with previous observations in other driving tasks that conventional temporal models provide only limited benefits in driving-prediction tasks[60,61]. This contrasts with human driving, in which current control decisions depend strongly on recent sensory and motion history[10]. This discrepancy suggests that simply introducing an explicit recurrent state is insufficient to reproduce the selective use of temporal context characteristic of biological working memory. Instead, the temporal properties of the memory mechanism, including how long information is retained and how obsolete information is discarded, may determine whether historical context facilitates or interferes with steering prediction.

To test whether the advantage of IPNet was associated with constrained retention rather than increased memory capacity, we systematically varied the amount of input history available to each model. The ResNet-LSTM exhibited a U-shaped dependence of RMSE on history length across both ResNet18 and ResNet34 backbones and under both daytime and nighttime conditions (Fig. 5d,f). A moderate history initially improved prediction, whereas further extension progressively increased the error. A similar degradation under extended temporal context was observed for the Video Swin Transformer (Fig. S12), indicating that this behavior was not restricted to recurrent architectures. These results show that access to a longer history is not uniformly beneficial in continuous driving streams, where temporally remote events may no longer be relevant to the current control state.

In contrast, IPNet maintained comparatively stable prediction errors as the observed sequence was extended (Fig. 5e,g). Together with the preceding finding that excessively slow intrinsic-state decay impaired temporal-order discrimination, this result supports an interpretation in which the finite decay of the MTJ state limits the accumulation of obsolete activity while preserving recent task-relevant transitions. Thus, the same capacity limitation and temporal forgetting that produced working-memory-like signatures in the controlled cognitive tasks provide a functional advantage in a substantially more complex continuous-control setting. The benefit of IPNet therefore arises not from retaining a more complete history, but from physically constraining which parts of that history continue to influence subsequent computation.

We further examined whether the position of intrinsic memory within the network affected steering prediction. Consistent with the Time-Reversed DVS Gesture results, performance was highest when the IP layer was placed directly at the sensory input (Fig. 5c). Moving the IP layer after convolutional feature extraction or towards the network output progressively increased RMSE and reduced EVA. A similar placement dependence was also observed in the n-back task (Fig. S7), suggesting that the effect is not restricted to a specific visual task. These results indicate that intrinsic temporal processing is most effective before fine-grained input dynamics are compressed by subsequent feature abstraction. Thus, placing computation at or near the sensor should not be viewed solely as a strategy for reducing data movement, memory traffic and latency. For in-sensor and near-sensor neuromorphic computing, early temporal processing can also represent a performance-driven architectural choice, because task-relevant temporal information is most accessible at the sensory frontier. Conceptually, this design principle parallels biological sensory systems, such as the retina and cochlea, where substantial temporal filtering and feature preprocessing occur at the sensory periphery before information is transmitted to downstream neural circuits.

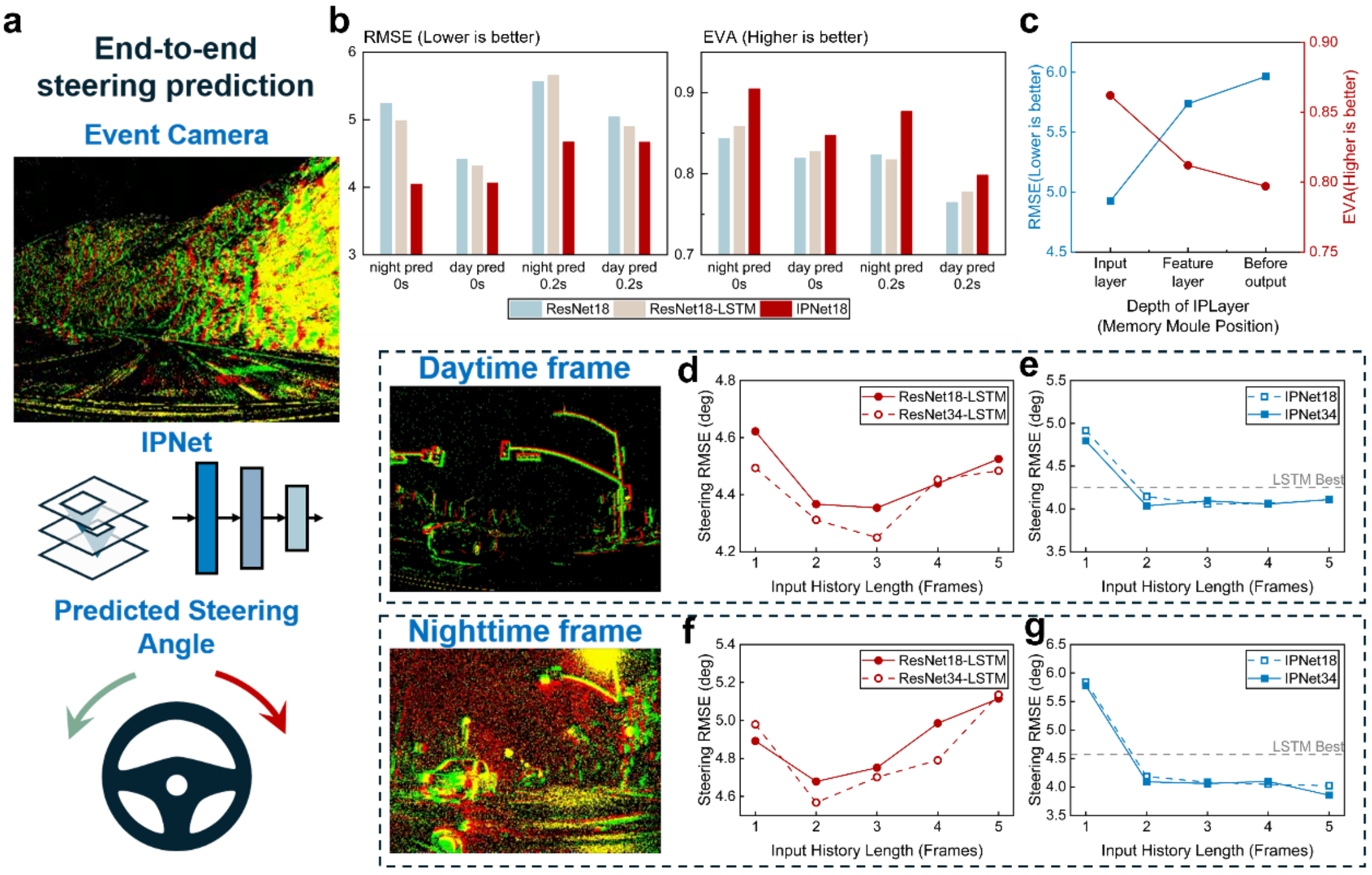

**Fig. 5. Real-world application for working memory: End-to-end steering prediction on the DDD20 dataset. a**, Schematic illustration of the end-to-end steering angle prediction framework. **b**, Performance evaluation under varying environmental conditions (day versus night) and prediction horizons (instantaneous steering angle versus 0.2 s look-ahead). The IPNet18 (red bars) is benchmarked against baselines using Root Mean Square Error (RMSE, lower is better) and Explained Variance Accuracy (EVA, higher is better). **c**, Impact of the memory module's depth on regression performance. RMSE (blue, left axis) and EVA (red, right axis) are plotted as a function of the IPLayer's position, illustrating the sensitivity of the task to the placement of temporal processing units. **d-g**, Impact of input history length on steering prediction performance across illumination conditions. Quantitative analysis of steering error (RMSE) as a function of input history length across distinct backbones

(ResNet18/34) reveals divergent temporal behaviors. ResNet-LSTM models (**d, f**) exhibit a U-shaped error profile regardless of the backbone, indicating that retaining excessive history becomes a liability due to sensory redundancy. In contrast, IPNet (**e, g**) demonstrates biological-like resilience, maintaining stable precision as historical context expands.

## Hardware-in-the-loop validation and robustness to device variability

To determine whether the working-memory-like dynamics predicted by the experimentally calibrated model can be retained when the intrinsic-memory computation is executed by physical devices, we implemented a hardware-in-the-loop system for the DDD20 steering-prediction task (Fig. 6a). The core IP layer was implemented using eight fabricated MTJ devices, comprising seven thermally sensitive devices and one thermally insensitive reference device. Through time multiplexing, these devices formed a 49-neuron IP layer operating with a 230-ns cycle, while the remaining network layers were executed in software (see Methods). Importantly, the temporal state of the hardware IP layer was carried directly by residual Joule heating generated by preceding input pulses, without introducing a separate software-level history variable to reproduce the intrinsic memory dynamics.

On the representative continuous driving sequence shown in Fig. 6b, the hardware-in-the-loop implementation closely reproduced the steering predictions of the experimentally calibrated software model. The RMSE increased only modestly from 4.020 in simulation to 4.198 in hardware, while the explained variance decreased from 0.953 to 0.948 (Fig. 6c), indicating measurable but limited discrepancies arising from device-level non-idealities. For a controlled comparison, the hardware-executed IP layer and the LSTM were inserted at the same position immediately after the ResNet feature extractor and were matched in state dimensionality. Under these architectural constraints, IPNet, whose temporal memory was implemented entirely by the physical states of the MTJ devices, still outperformed the matched full-precision software LSTM, which yielded an RMSE of 5.77 and an explained variance of 0.939 on the same sequence. This result indicates that the performance advantage cannot be attributed simply to differences in memory-module placement or size, and that the capacity-limited intrinsic dynamics retained their task-level benefit when the temporal-memory computation was executed in physical hardware.

These results establish two distinct aspects of physical validation. First, the close correspondence between the hardware and simulated outputs supports the fidelity of the device-calibrated intrinsic-plasticity model. Second, the superior performance relative to the LSTM baseline indicates that the task-level advantage of the capacity-limited and decaying intrinsic state is retained when the core temporal computation is performed by physical MTJ devices. The experiment therefore goes beyond reproducing an isolated device response: it shows that Joule-heating-mediated intrinsic dynamics can participate directly in continuous, task-relevant temporal processing.

We further assessed the robustness of IPNet to device-to-device variability using the experimentally characterized parameter distributions of the 200-nm-diameter MTJs employed in the hardware-in-the-loop experiment (Supplementary Table 1). The nominal IPNet was trained using the parameters extracted from a single device, whereas, during inference, each neuron was independently assigned parameters sampled from the measured distributions. Across five random device-array realizations, the unadapted network remained robust to moderate variability, retaining an accuracy of 99.10 ± 0.08% when the parameter standard deviations were scaled to 50% of their experimentally characterized values. Performance declined more rapidly at larger variability levels, reaching 82.33 ± 0.38% at 80%, 62.53 ± 0.96% at 100% and 21.56 ± 0.63% at 200% (Fig. S15). Device-aware fine-tuning largely compensated for this degradation, restoring accuracy to 99.48%, 99.83% and 98.61% at variability levels of 100%, 150% and 200%, respectively. These results indicate that IPNet tolerates moderate device variability without retraining, while substantially larger variations can be accommodated through device-aware adaptation.

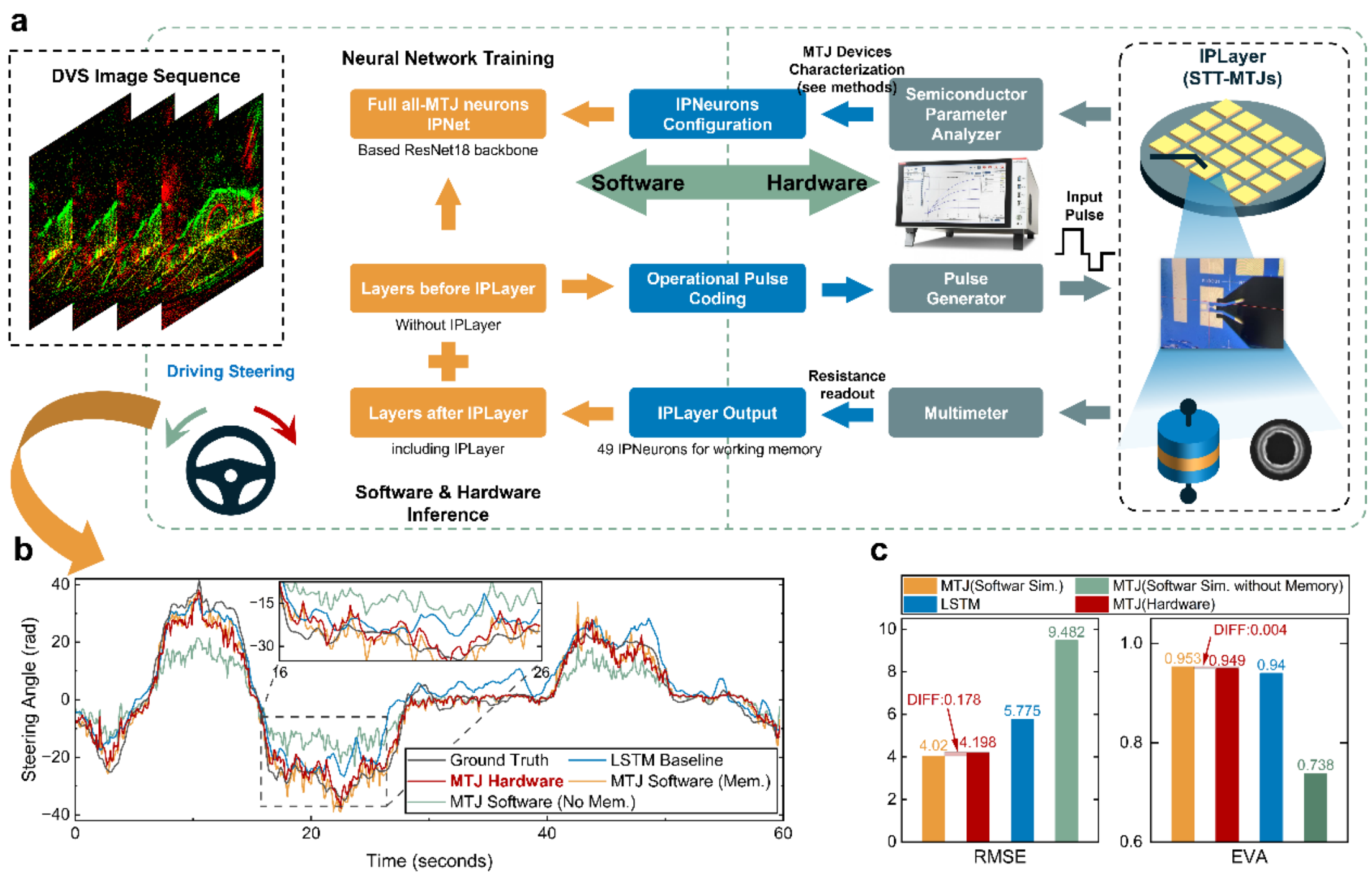

**Fig. 6. Hardware-in-the-loop validation of MTJ-based intrinsic plasticity enabled**

**working memory. a**, Schematic illustration of the hardware-in-the-loop (HIL) experimental framework applied to the DDD20 end-to-end steering prediction task. **b**, Representative steering angle trajectories over a continuous 60-second (600-frame) test sequence. **c**, Quantitative performance characterization of the hardware implementation, software simulation, and LSTM baseline over the 600-frame test sequence shown in Fig. 6b. The results demonstrate a high degree of consistency between the physical MTJ hardware and software models (with marginal deviations of 0.178 in RMSE and 0.004 in EVA), confirming the reliability of the simulation. Furthermore, the MTJ-based hardware system outperforms the full-precision software LSTM, validating the physical realization of efficient, human-like working memory advantages.

## Discussion

This work shows that neuronal intrinsic plasticity alone can generate selected signatures associated with human working memory and that the same constrained temporal dynamics can benefit dynamic visual processing. IPNet, which exhibited these human-like working-memory signatures, outperformed conventional recurrent, spatiotemporal convolutional and attention-based architectures on temporally sensitive dynamic-vision tasks. On the Time-Reversed DVS Gesture task, IPNet achieved an accuracy of 99.83%, compared with 95.31% for ResNet-LSTM, 96.88% for R(2+1)D and 96.35% for the pretrained Video Swin Transformer. This advantage extended to continuous steering prediction on DDD20, where IPNet reduced the mean RMSE by 12.4% relative to the matched LSTM baseline. Crucially, while traditional memory architecture like LSTM and Transformer exhibited a U-shaped error profile—deteriorating due to historical noise accumulation when forced to retain excessive context—IPNet maintained temporal robustness by utilizing bio-mimetic decay as an optimal filter. This system-level dominance translates not only into superior performance but also into profound energy efficiency gains, offering a stark contrast to traditional GPU-based systems. In dynamic vision tasks, IPNet reduces the working memory energy overhead by 2,874× compared to LSTMs and by 90,920× compared to 3D-CNNs, while maintaining superior accuracy (Fig. S11). Together, these results indicate that device-intrinsic dynamics can implement selected functional constraints associated with biological working memory and translate them into measurable advantages in temporal processing and energy efficiency over the conventional architectures examined here.

Mechanistically, IPNet realizes working-memory-like functions through a neuron-local intrinsic state, without recurrent connectivity, synapse-specific short-term memory variables or explicit network-level history buffers. By repurposing Joule heating—conventionally treated as a parasitic effect—as an activity-dependent computational state, recent inputs transiently alter the switching probability of MTJ neurons and thereby modulate their responses to subsequent inputs before the state relaxes towards equilibrium. Information is thus retained as a latent change in neuronal responsiveness rather than through persistent spiking activity or an explicitly stored digital history, providing an engineered analogue of activity-silent working memory[6,7,20]. These results demonstrate that working memory functions can emerge from the intrinsic dynamics of artificial neurons themselves, shifting temporal memory from a separately engineered network-level module to a physical state embedded within neuron-level device.

Architecturally, the "Memory-at-the-Frontier" effect identifies where constrained intrinsic memory is most effective within a sensory-processing hierarchy. Across the n-back, Time-Reversed DVS Gesture and steering-prediction tasks, performance was consistently highest when the IP layer was placed close to the input and deteriorated as it was moved towards deeper representations. This placement dependence suggests that the intrinsic state is most useful while fine-grained temporal transitions remain directly accessible in the sensory stream, before they are compressed or discarded by successive stages of spatial feature abstraction. Near-sensor temporal processing may therefore provide not only conventional system-level benefits, such as reduced data movement and latency, but also a task-level computational advantage. This functional organization is analogous to biological sensory systems, including the retina and cochlea, where substantial temporal adaptation and filtering occur at the sensory periphery before downstream processing[62–64].

Finally, the computational principle established here is not necessarily restricted to MTJs or spiking neural networks. Its essential requirements are an activity-dependent internal state that can modulate subsequent responses, relax over a task-relevant timescale and be integrated into neural computation without a separately maintained history variable. Other volatile or dissipative device platforms, including phase-change[65], resistive memories[66] and ferroelectric devices[67], may potentially support related forms of intrinsic temporal processing, although their suitability will depend on state controllability, variability, peripheral-circuit requirements and the correspondence between device and task timescales. Importantly, this approach does not require highly specialized fabrication designs and remains compatible with existing CMOS processes. At the algorithmic level, the improvement obtained by incorporating the IP layer into ConvNeXt V2 and Swin Transformer indicates that this temporal-processing mechanism can also operate in non-spiking networks, rather than being specific to SNN dynamics. Together with the compact estimated neuron footprint of approximately 1.5 $\mu m^2$, these results position device-intrinsic relaxation as a potentially general computing primitive for implementing finite and adaptive temporal context. More broadly, the present work suggests that the value of device physics lies not only in reducing the cost of an existing digital memory operation, but also in embedding biologically inspired constraints, particularly finite retention and selective forgetting, directly into the computational substrate.

## Methods

### MTJ fabrication

The MTJ devices are fabricated with a stack structure as follows, from the substrate side:Ta(5)/CuN(20)/Ta(5)/Pt(5)/$[Co(0.4)/Pt(0.6)]_5$/Ru(0.85)/$[Co(0.4)/Pt(0.6)]_3$/Ta(0.5) /$Co_{20}Fe_{60}B_{20}$(0.9)/MgO(0.9)/$Co_{20}Fe_{60}B_{20}$(1.1)/Ta(0.5)/$[Co(0.3)/Pt(1.5)]_2$/Ru(5) (Fig. 2a). The numbers in parentheses indicate the layer thickness in nanometers. The films were patterned into nanopillars using a standard nanofabrication process. First, the bottom electrodes were defined by ultraviolet (UV) photolithography and ion beam etching. Next, the MTJ pillars were patterned using electron-beam lithography (EBL) and etching. An insulating $SiO_2$ layer was then deposited, followed by a lift-off process to expose the pillar tops. Finally, the top contact pads were fabricated using UV photolithography followed by metal deposition and a lift-off process.

**Modeling of MTJ-based thermal intrinsic plasticity**

We first characterize a simple static MTJ-based stochastic neuron (Fig. 2d), which also serves as the fundamental model for the non-plastic neurons in the SNN-based IPNet. For a voltage divider circuit comprising a Magnetic Tunnel Junction (MTJ) and a reference resistor, the input-output relationship is defined as:

$$V_{OUT} = \frac{R_{MTJ}}{R_{MTJ} + R_{ref}} V_{IN} \tag{1}$$

where $V_{OUT}$ and $V_{IN}$ denote the output and input voltages, and $R_{MTJ}$ and $R_{ref}$ represent the resistances of the MTJ and the reference resistor, respectively. The output $V_{OUT}$ drives a comparator with a threshold voltage $V_{thr}$. To ensure correct state detection, $R_{MTJ}$ and $R_{ref}$ must satisfy:

$$\frac{R_P}{R_P + R_{ref}} V_{READ} < V_{thr} < \frac{R_{AP}}{R_{AP} + R_{ref}} V_{READ} \tag{2}$$

where $V_{READ}$ is the read voltage, and $R_P$ and $R_{AP}$ are the MTJ resistances in the parallel (P) and anti-parallel (AP) states, respectively. During a read operation ($V_{IN} = V_{READ}$), Eqs. (1) and (2) ensure that $V_{OUT} > V_{thr}$ (firing) when the MTJ is in the AP state, and $V_{OUT} < V_{thr}$ (non-firing) when in the P state. Thus, the neuron's firing state is determined exclusively by the MTJ configuration. Assuming the MTJ is reset to the P state at the onset of each time step, the neuronal firing probability corresponds to the probability of the MTJ switching from P to AP (Fig. 2e), given by:

$$P_{fire} = P_{sw}(V_{MTJ}) = \frac{L}{1 + e^{-k(V_{MTJ} - x_0)}} \quad (3)$$

where $P_{fire}$ is the firing probability, $P_{sw}$ is the switching probability, and $V_{MTJ}$ is the voltage across the MTJ. $L$ denotes the maximum switching probability for the given timing, while $k$ and $x_0$ represent the slope and midpoint of the sigmoid function, respectively.

Eq. (3) describes the static stochastic response. We next extend this to model a dynamic MTJ stochastic neuron with intrinsic plasticity (IP). The state modulation induced by Joule heating is parameterized by the shift in the sigmoid midpoint $x_0$ (Fig. 2f, g). Consequently, the firing probability $P_{fire}(t)$ at time $t$ is expressed as:

$$P_{fire}(t) = \frac{L}{1 + e^{-k\left(V_{MTJ}(t) - x_0(t)\right)}} \quad (4)$$

where $x_0(t)$ and $V_{MTJ}(t)$ are the instantaneous MTJ state and input voltage, respectively. During operation, the evolution of $x_0(t)$ is governed by the competition between Joule heating (driven by $V_{MTJ}(t)$) and natural thermal dissipation. We define $x_0(t)$ as the deviation from an ambient baseline:

$$x_0(t) = x_{0,env} - \Theta(t) \quad (5)$$

$$\tau_{the} \frac{d\Theta(t)}{dt} = -\Theta(t) + s\left(V_{MTJ}(t)\right) \quad (6)$$

where $\Theta(t)$ is the state offset at time $t$, and $\tau_{the}$ is the thermal time constant. Eq. (6) presents the differential dynamics of the IP neuron, where state evolution depends on the current offset $\Theta(t)$ and the input $V_{MTJ}(t)$. The term $S(\cdot)$ defines the coupling between the state offset and the input voltage. The right-hand side of Eq. (6) separates the contributions of thermal dissipation (first term) and Joule heating (second term). Approximating these as independent processes over a time interval $\Delta t$ yields the discrete dynamics for cooling and heating:

$$\{\Theta_t - \Theta_{t-1}\}_{cooling} = \Theta_{t-1}\left(e^{-\frac{\Delta t}{\tau_{the}}} - 1\right) \tag{7}$$

$$\{\Theta_t - \Theta_{t-1}\}_{heating} = \frac{1}{\tau_{the}}\int_{t-\Delta t}^{t} s\left(V_{MTJ}(t)\right) dt \tag{8}$$

Eqs. (7) and (8) represent the discrete cooling and heating components, respectively. When $\Delta t$ equals the neural network time step $T_{pulse}$, the heating integral becomes independent of the specific start time $t$ and is denoted as $S(\cdot)$. The equations then simplify to:

$$\{\Theta_t - \Theta_{t-1}\}_{cooling} = \Theta_{t-1}(\gamma - 1) \tag{9}$$

$$\{\Theta_t - \Theta_{t-1}\}_{heating} = \frac{1}{\tau_{the}} S\left(V_{MTJ}, T_{pulse}\right) \tag{10}$$

As shown in Fig. S2, Eq. (9) demonstrates excellent agreement with experimental data under cooling-only conditions (i.e., no input pulse). Furthermore, by characterizing the state offset in the subsequent time step ($\Theta_t$) as a function of input voltage ($V_{MTJ}$) from an initial zero-offset state ($\Theta_{t-1} = 0$), we derive:

$$\{\Theta_t - \Theta_{t-1}\}_{heating} = \alpha V_{MTJ}^2 + \beta V_{MTJ} \tag{11}$$

where $\alpha$ and $\beta$ are constant coefficients. Combining Eqs. (9) and (11) provides the complete discrete dynamical equation for the IP neuron:

$$\Theta_t = \gamma \Theta_{t-1} + \alpha V_{MTJ}^2 + \beta V_{MTJ} \tag{12}$$

By integrating Eqs. (4), (5), and (12), the thermally plastic MTJ neuron firing probability can be calculated for any given time and state. For a thermally insensitive MTJ, where both $\alpha$ and $\beta$ are zero, the state remains time-invariant and equals $x_{0,env}$, as follows:

$$\Theta_0 = x_{0,env} - x_0(t=0) = x_{0,env} - x_{0,env} = 0 \tag{13}$$

$$\Theta_t = \Theta_{t-1} = \cdots = \Theta_0 = 0 \tag{14}$$

$$x_0(t) = x_{0,env} - \Theta(t) = x_{0,env} \tag{15}$$

The software baseline shown in Fig. 6b, simulated using this formulation (Eqs. (4), (5), (12) and (15)), aligns closely with the hardware experimental results, validating the accuracy of the proposed IP neuron model.

**Device characterization**

All electrical measurements were performed using a Keithley 4200A-SCS semiconductor parameter analyzer equipped with a 4225-PMU pulse measure unit and a 4225-RPM remote amplifier/switch. All measurements were performed at room temperature and atmospheric pressure in the absence of an external magnetic field. The switching probability $P_{sw}$ for each data point was averaged from over 10,000 repeated cycles.

First, the baseline voltage-dependent switching probability was measured and fitted to Eq. (3) (Fig. 2e). To characterize thermal sensitivity, we employed a pump-probe scheme: a pre-heating pulse (0.9 V, 50–500 ns) was applied, followed by a 20 ns RESET pulse to initialize the MTJ device to the P state, after which the switching probability was immediately measured. Thermal relaxation dynamics were subsequently characterized by introducing a variable cooling delay following a fixed pre-heating pulse (0.9 V, 500 ns) (Fig. 2g). Data across all thermal states used the sigmoid midpoint $x_0$ as the sole state variable representing device temperature, while keeping other parameters constant.

To quantify voltage-dependent heating effects, the device was subjected to pre-heating pulses of varying amplitudes with a fixed plateau duration of 150 ns (rise/fall times fixed at 20 ns by the instrument limit). The resulting thermal state was immediately probed using a standard probe pulse (0.83 V, 150 ns). The induced state shift, $\Theta_t$, was extracted and fitted as a function of the pre-pulse voltage according to Eq. (11) (Fig. S1). Similarly, cooling dynamics were quantified by heating the device to a saturated temperature using a train of five pre-heating pulses (each period comprising: 20 ns rise, 0.93 V/150 ns Set, 20 ns fall, -0.8 V/20 ns Reset, and 20 ns delay), followed by variable cooling intervals (0–5000 ns). The decay of $\Theta(t)$ over time was fitted using Eq. (9) (Fig. S2). The accuracy of the derived thermal model was validated by comparing the predicted state evolution with experimental results from a sequence of three consecutive pulses (0.83 V, 150 ns each), showing excellent agreement (Fig. S4). This validation is further corroborated by the hardware-in-the-loop (HIL) experimental results presented in Fig. 6b. These hardware-extracted state evolution functions and parameters were utilized for all MTJ-based neural network simulations reported in this work.

**MTJ-based neuron circuit design and area estimation**

Our artificial neuron is built upon a standard magnetic tunnel junction (MTJ) (Fig. 2a), comprising a free magnetic layer and a pinned magnetic layer separated by an oxide tunnel barrier. Free layer magnetization can be manipulated by external stimuli, while pinned layer magnetization remains fixed. The device state is defined by the orientation of magnetization direction of the two magnetic layers. The low (high) resistance parallel (anti-parallel) state refers to the case where the two magnetizations are in the same (opposite) direction. Upon current injection, together with thermal fluctuations, the spin-transfer torque (STT) induces probabilistic magnetization switching of the free layer (Fig. 2b).

A fundamental MTJ-based stochastic neuron circuit is illustrated in Fig. 2d. To enhance temporal processing capabilities and robustness, we implemented a differential neuron architecture (Fig. S8). This design comprises two elementary neurons: a "temporal" neuron incorporating a thermally sensitive MTJ (e.g., 200 nm diameter) and a "reference" neuron incorporating a thermally insensitive MTJ (e.g., 100 nm diameter, Fig. S9).

While both devices share the same baseline switching probability, their dynamic behaviors differ. The outputs, denoted as $Spike_{the}$ and $Spike_{ref}$, are stochastic variables generated via Bernoulli sampling based on the firing probability derived in the previous section: $Spike \sim Bernoulli(P_{fire})$. Specifically, $Spike_{the}$ is governed by the full thermal dynamics (Eqs. 4–12), whereas $Spike_{ref}$ corresponds to the time-invariant case where $\alpha = \beta = 0$ (Eqs. 13–15).

The differential unit provides two output channels: $Spike_{ref}$, representing the instantaneous input intensity, and the difference signal $Spike_{diff} = Spike_{the} - Spike_{ref}$, encoding historical thermal information. This operation is implemented using a subtractor circuit connecting the two neurons. To minimize the circuit footprint, the voltage-dividing resistors shown in Fig. 2d were replaced by MTJ devices of equivalent resistance, and the comparator was implemented using two series-connected CMOS inverters[68]. In the system architecture, a single input feeds into $N$ identical differential neurons. The outputs are average-pooled before transmission to the synapse, effectively performing $N$ parallel Monte Carlo simulations within a single time step. All simulations and energy and area estimates for the IP layer are based on this differential-neuron configuration, whereas neurons in the subsequent layers use the standard non-differential circuit shown in Fig. 2d.

Given the negligible footprint of BEOL-integrated MTJs compared to the CMOS logic, the circuit area is determined by the transistor count. Reported compact memory-cell layouts typically occupy $60–240F^2$, depending on the cell architecture and technology node.[69–71] Here, we conservatively assume an effective footprint of $80F^2$ per transistor, including layout and routing overhead. Accordingly, the 24-transistor differential neuron is estimated to occupy approximately $1.5\ \mu\text{m}^2$ at $F = 28$nm.

**Neural network training and evaluation**

All neural network training and evaluation were implemented in Python 3.9/3.10. Computations were performed using NVIDIA A800 or H100 GPUs.

**n-back.** All networks (IPNet, LSTM, and LIF-SNN implemented via SpikingJelly[72]) utilized a 225-225-2 topology. For the LSTM, the intermediate stage consisted of a single recurrent layer with 225 hidden units. Specifically, the IPNet and LIF-SNN incorporated neurons at the input layer, whereas the LSTM received inputs directly via synapses. The input images (225 pixels) consisted of a 3×3 grid of 5×5 pixel blocks, where a total of 8 distinct categories (each represents a letter from A to H) were encoded by randomly coloring 4 out of the 9 blocks. Data were input serially in sequences of 20 randomly sampled patterns. At each time step, the network determined if the current input matched the input presented ***n*** steps earlier. The performance was evaluated using the metric PR = TPR - FPR[42], excluding the first ***n*** steps of each sequence.

**Free recall.** The networks (IPNet and LSTM) utilized a fully connected architecture with a 225–512–50 topology. The LSTM consisted of a single recurrent layer with 512 hidden units. As in the N-back task, the IPNet incorporated neurons at the input layer, whereas the LSTM received inputs directly via synapses. The input images were encoded using the same block-based method as described above, but with a total of 50 distinct categories. For each trial, a sequence of 15 unique patterns sampled from the 50 categories was input serially. Following the presentation of the sequence and a variable delay period of $T$ time steps, the network was required to output the class indices of all 15 presented items simultaneously. In this free recall task, the network was only required to identify the presence of the items, independent of their temporal order. The recall accuracy was quantified as a function of the item's serial position during input and the duration of the delay period $T$.

**Proactive and retroactive interference.** The networks were configured with a 900-225-50 topology. Consistent with the previous tasks, the IPNet incorporated neurons at the input layer, whereas the LSTM comprised a single recurrent layer with 225 hidden units. To investigate the role of data similarity, inputs were generated using two distinct encoding schemes: one-hot encoding (representing orthogonal, non-overlapping inputs) and 4-out-of-9 coding (representing inputs with partial similarity). For both schemes, the input dimension was standardized to 900 units, representing 50 distinct categories. The experiment consisted of three conditions, each spanning 3 time-steps. In the control (no-interference) task, the target memory item was presented at the second time-step ($t$=2), with null inputs at $t$=1 and $t$=3. The network was trained to output the label of the target item at the end of the sequence. In the proactive interference task, a distractor item was presented at $t$=1, preceding the target item at $t$=2. In the retroactive interference task, the distractor was presented at $t$=3, following the target item at $t$=2. In all interference trials, the distractor and target were drawn from the same distribution. To emulate the protocol of analogous human cognitive tasks, the networks were trained to retain both the distractor and the target; however, the interference effect was characterized specifically by the recall accuracy of the target item.

**Cued recall.** The experimental configuration and network architecture remained identical to those of the uncued interference tasks. The 50 distinct categories were grouped into five superordinate classes, each comprising 10 subcategories. Input presentation was identical to the uncued tasks; however, during the recall phase, the network was provided with a cue specifying the superordinate class of the target memory item. This cue was implemented by highlighting all pixels corresponding to that superordinate class (Fig. S5), thereby restricting the network to retrieve the target exclusively from within this specific subset.

**Dynamic vision tasks.** To enhance the processing capability for complex visual stimuli, convolutional neural network (CNN)-based architectures were adopted as the backbone for both the proposed IPNet and the baseline models. For the IPNet architectures, we utilized fully spiking SEW-ResNet frameworks[52], where IPNet18 and IPNet34 were adapted from SEW-ResNet18 and SEW-ResNet34, respectively. All IPNet models (including FCN-IPNet in memory tasks) were trained directly using the Backpropagation Through Time (BPTT) algorithm. Unless otherwise specified, the input layer was replaced by the proposed differential neuron block (designated as the IPLayer). In contrast, all subsequent layers employed the thermally insensitive stochastic MTJ neuron model; these neurons are implemented without the differential architecture, strictly simulating the standard circuit configuration illustrated in Fig. 2d. Additionally, the original Batch Normalization (BatchNorm) layers in the SEW-ResNet were replaced with Root Mean Square Normalization (RMSNorm). Specific to the Time-Reversed DVS Gesture task, the IPLayer was configured to perform 8 parallel Monte Carlo simulations (via average pooling of eight parameter-identical neurons with independent Bernoulli sampling), whereas all downstream MTJ neurons operated with a single simulation count, a configuration found to yield optimal performance. For the baselines, the ResNet18-LSTM model integrated the standard PyTorch ResNet18 backbone with a single-layer LSTM containing 1024 hidden units, positioned between the backbone output and the final classification layer. The R(2+1)D-18 baseline utilized the standard implementation provided in the PyTorch library. To accommodate varying input spatial dimensions across different tasks, the first convolutional layer of all models was modified accordingly. To ensure a fair comparison, all models underwent equivalent hyperparameter optimization.

For ConvNeXt, we used a standard, non-spiking ConvNeXt V2-tiny model[54] as both the baseline and the IPNet backbone, initialized with ImageNet-22k pre-trained weights. In the IPNet implementation, images were sequentially input into the MTJ IPLayer to generate pulses. These pulses were then converted to floating-point numbers via rate coding before being fed into the non-spiking ANN backbone. We evaluated this model configuration on the DailyDVS and HARDVS datasets, utilizing exclusively DVS data as the input modality for both.

For Transformer, we integrated the IPNet using an approach similar to the ConvNeXt implementation. An IPLayer was inserted before a standard Swin Transformer-tiny backbone[73] to process temporal information. The generated pulses were then encoded into floating-point numbers and fed into the ANN. This backbone was initialized with ImageNet-1K pre-trained weights. We utilized a Video Swin Transformer-tiny[15] as the spatiotemporal baseline, initialized with Kinetics-400 pre-trained weights. All input images for this model were resized to 224 × 224 pixels prior to processing by the Transformer models.

**Experimental settings for hardware-in-the-loop (HIL)**

**Network architecture and hardware mapping.** To facilitate physical implementation, a customized IPNet architecture was developed. The convolutional backbone remains identical to the IPNet18; however, the thermally sensitive IPLayer was relocated from the input stage to the post-convolution stage. Specifically, the feature maps from the convolutional backbone are projected via a fully connected (FC) layer to an IPLayer comprising 49 neurons, which is then connected to the final steering-output layer via another FC layer. To map these 49 logical neurons into limited hardware resources, we employed a time-division multiplexing strategy: 7 physical thermally sensitive MTJ devices were each reused 7 times. A single thermally insensitive MTJ served as the shared reference device for the differential neuron architecture. During both training and inference, the network processes a sequence of 4 consecutive video frames serially for each steering angle prediction, ensuring that the thermal history within the MTJs is effectively utilized for temporal processing. For each logical neuron, the complete four-frame sequence was applied consecutively to preserve its thermal history. Before the same physical MTJ was reassigned to another logical neuron, the device was allowed to return to its ambient thermal state. For comparison, the ResNet18-LSTM baseline was adapted with a similar topology, where the LSTM hidden size was reduced from 1024 to 49 units to match the scale of the hardware IPNet.

**HIL execution protocol.** First, the network was trained entirely in a Python software environment to obtain optimal synaptic weights. Subsequently, inference was performed on the test dataset, which comprised two distinct 600-frame video segments totaling 1,200 samples (results visualized in Fig. 6b and Fig. S10), to extract the precise voltage inputs destined for the IPLayer at each time step. These voltage values were encoded into pulse sequences and applied to the corresponding physical MTJ devices (both the sensitive and reference units) via radio-frequency (RF) probes, preserving the exact temporal order of the input frames. Crucially, identical input sequences were applied to both the thermally sensitive and reference MTJs. This setup aligned with the software training protocol, where the reference neuron was explicitly modeled to exhibit a switching probability independent of historical pulses, thereby experimentally validating that the fabricated devices can support both memory-dependent and memory-independent modes under identical stimulation. The switching state (Parallel/Anti-Parallel) of the MTJs was measured to determine the spiking output. To emulate the average pooling mechanism used in the software model, 32 repeated Monte Carlo simulations were conducted for each neuron per frame on the hardware. Finally, the experimentally obtained average firing rates were fed back into the software environment to compute the final steering angle through the last fully connected layer.

**Working memory energy measurement**

**Experimental IPNet energy estimation.** To provide an estimated comparison with conventional GPU-based temporal processing, the energy consumption of the IPNet was quantified using a physics-based projection derived from direct experimental measurements. We calculated the energy expenditure using the applied pulse amplitude ($V_{amp} = 0.80V$) and the device resistance derived from I-V curves. Although the effective voltage across the MTJ is lower due to parasitic voltage drops in the experimental apparatus (e.g., probes and cables), we utilized the source voltage to establish a conservative upper bound for the energy consumption per operation. For a representative thermally sensitive MTJ (ID: R09C11), the resistance states were measured as $R_P = 230\Omega$ and $R_{AP} = 410\Omega$. Under a standard operating voltage of 0.80 V (yielding a baseline switching probability of ≈19.72%) and a total pulse protocol duration of 230 ns (encompassing both Set and Reset phases), the energy consumption per cycle is approximately 450 pJ. The thermally insensitive reference MTJ, designed with higher resistance, consumes approximately 82 pJ per cycle under identical conditions. The power dissipation of the auxiliary CMOS circuitry (24 transistors) in the differential neuron is negligible compared to the MTJ write energy. Accordingly, based on the HIL-derived MTJ operating parameters and allowing for the comparatively small CMOS overhead, the energy consumption of a single differential-neuron operation was conservatively rounded up to approximately 0.6 nJ. To project the estimated memory-related IP layer energy for the DVS Gesture task (2×128×128 resolution, 50 frames per sample), we considered an average input spike density of 8.9%. This yields an estimated memory energy consumption of 87 uJ (2×128×128×8.9%×50×0.6 nJ) per sample for the IPNet. While optimal performance in general tasks may require 4-16 parallel Monte Carlo simulations (increasing consumption to 348-1392 uJ), it is notable that for the specific Time-Reversed DVS Gesture task, a single simulation count is sufficient to outperform the ResNet-LSTM baseline, maintaining the minimal energy footprint (Fig. S11).

**Baseline model energy measurement.** To quantify the specific energy cost attributed to memory processing in conventional ANN memory architectures, we measured the incremental power consumption relative to a static backbone. Experiments were conducted on an NVIDIA A800 GPU (fabricated on an advanced TSMC process node). During inference, power consumption was monitored using pynvml (NVIDIA's own energy reporting framework). We compared the inference energy of the temporal models (ResNet18-RNN, ResNet18-LSTM, and (2+1)D ResNet18) against a standard static ResNet18 baseline, with all models utilizing identical batch sizes and hyperparameters. Energy values represent the median over 20 iterations of the Time-Reversed DVS Gesture test set. The incremental energy costs per sample, attributed to the memory/temporal components, were measured as 0.21 J for the RNN, 0.25 J for the LSTM, and 7.91 J for the (2+1)D ResNet18. Comparative results indicate that the proposed IPNet architecture reduces incremental temporal-processing energy consumption by 100-2,800× of magnitude relative to standard recurrent baselines (RNN/LSTM), extending to approximately 5,000-90,000× orders of magnitude when compared against computationally intensive spatiotemporal models such as the (2+1)D ResNet.

## References

1. Constantinidis, C. & Klingberg, T. The neuroscience of working memory capacity and training. *Nat Rev Neurosci* **17**, 438–449 (2016).
2. Olivers, C. N. L., Peters, J., Houtkamp, R. & Roelfsema, P. R. Different states in visual working memory: when it guides attention and when it does not. *Trends in Cognitive Sciences* **15**, 327–334 (2011).
3. Brady, T. F., Robinson, M. M. & Williams, J. R. Noisy and hierarchical visual memory

across timescales. *Nat Rev Psychol* **3**, 147–163 (2024).

4. Egorov, A. V., Hamam, B. N., Fransén, E., Hasselmo, M. E. & Alonso, A. A. Graded persistent activity in entorhinal cortex neurons. *Nature* **420**, 173–178 (2002).
5. Fitz, H. *et al.* Neuronal spike-rate adaptation supports working memory in language processing. *Proceedings of the National Academy of Sciences* **117**, 20881–20889 (2020).
6. Lundqvist, M., Herman, P. & Miller, E. K. Working memory: delay activity, yes! Persistent activity? Maybe not. *Journal of neuroscience* **38**, 7013–7019 (2018).
7. Panichello, M. F. *et al.* Intermittent rate coding and cue-specific ensembles support working memory. *Nature* **636**, 422–429 (2024).
8. Mongillo, G., Barak, O. & Tsodyks, M. Synaptic theory of working memory. *Science* **319**, 1543–1546 (2008).
9. Debanne, D., Inglebert, Y. & Russier, M. Plasticity of intrinsic neuronal excitability. *Current opinion in neurobiology* **54**, 73–82 (2019).
10. Land, M. F. & Lee, D. N. Where we look when we steer. *Nature* **369**, 742–744 (1994).
11. Kiani, R., Hanks, T. D. & Shadlen, M. N. Bounded integration in parietal cortex underlies decisions even when viewing duration is dictated by the environment. *Journal of Neuroscience* **28**, 3017–3029 (2008).
12. Hochreiter, S. & Schmidhuber, J. Long short-term memory. *Neural computation* **9**, 1735–1780 (1997).
13. Tran, D. *et al.* A closer look at spatiotemporal convolutions for action recognition. in *Proceedings of the IEEE conference on Computer Vision and Pattern Recognition* 6450–6459 (2018).

14. Vaswani, A. *et al.* Attention is All you Need. in *Advances in Neural Information Processing Systems* vol. 30 (Curran Associates, Inc., 2017).

15. Liu, Z. *et al.* Video swin transformer. in *Proceedings of the IEEE/CVF conference on computer vision and pattern recognition* 3202–3211 (2022).

16. Sze, V., Chen, Y.-H., Yang, T.-J. & Emer, J. S. Efficient processing of deep neural networks: A tutorial and survey. *Proceedings of the IEEE* **105**, 2295–2329 (2017).

17. Zeng, A., Chen, M., Zhang, L. & Xu, Q. Are transformers effective for time series forecasting? in *Proceedings of the AAAI conference on artificial intelligence* vol. 37 11121–11128 (2023).

18. Hassabis, D., Kumaran, D., Summerfield, C. & Botvinick, M. Neuroscience-inspired artificial intelligence. *Neuron* **95**, 245–258 (2017).

19. Nazeri, A. & Pisu, P. LSTM-based Load Forecasting Robustness Against Noise Injection Attack in Microgrid. *arXiv preprint arXiv:2304.13104* (2023).

20. Stokes, M. G. 'Activity-silent'working memory in prefrontal cortex: a dynamic coding framework. *Trends in cognitive sciences* **19**, 394–405 (2015).

21. Rose, N. S. *et al.* Reactivation of latent working memories with transcranial magnetic stimulation. *Science* **354**, 1136–1139 (2016).

22. Wolff, M. J., Jochim, J., Akyürek, E. G. & Stokes, M. G. Dynamic hidden states underlying working-memory-guided behavior. *Nature neuroscience* **20**, 864–871 (2017).

23. Kozachkov, L. *et al.* Robust and brain-like working memory through short-term synaptic plasticity. *PLoS computational biology* **18**, e1010776 (2022).

24. Pozzorini, C., Naud, R., Mensi, S. & Gerstner, W. Temporal whitening by power-law

adaptation in neocortical neurons. *Nature neuroscience* **16**, 942–948 (2013).

25. Tavanaei, A., Ghodrati, M., Kheradpisheh, S. R., Masquelier, T. & Maida, A. Deep learning in spiking neural networks. *Neural networks* **111**, 47–63 (2019).

26. Delbruck, T. Frame-free dynamic digital vision. in *Proceedings of Intl. Symp. on Secure-Life Electronics, Advanced Electronics for Quality Life and Society* vol. 1 21–26 (Tokyo, 2008).

27. Liao, F., Zhou, F. & Chai, Y. Neuromorphic vision sensors: Principle, progress and perspectives. *Journal of Semiconductors* **42**, 013105 (2021).

28. Bellec, G., Salaj, D., Subramoney, A., Legenstein, R. & Maass, W. Long short-term memory and learning-to-learn in networks of spiking neurons. *Advances in neural information processing systems* **31**, (2018).

29. Yao, M. *et al.* Spike-driven transformer. *Advances in neural information processing systems* **36**, 64043–64058 (2023).

30. Indiveri, G. Neuromorphic is dead. Long live neuromorphic. *Neuron* **113**, 3311–3314 (2025).

31. Roy, K., Jaiswal, A. & Panda, P. Towards spike-based machine intelligence with neuromorphic computing. *Nature* **575**, 607–617 (2019).

32. Shaban, A., Bezugam, S. S. & Suri, M. An adaptive threshold neuron for recurrent spiking neural networks with nanodevice hardware implementation. *Nature Communications* **12**, 4234 (2021).

33. Yuan, R. *et al.* A neuromorphic physiological signal processing system based on VO2 memristor for next-generation human-machine interface. *Nature Communications* **14**,

3695 (2023).

34. Marković, D., Mizrahi, A., Querlioz, D. & Grollier, J. Physics for neuromorphic computing. *Nature Reviews Physics* **2**, 499–510 (2020).

35. Chatterjee, S., Salahuddin, S., Kumar, S. & Mukhopadhyay, S. Impact of self-heating on reliability of a spin-torque-transfer RAM cell. *IEEE transactions on electron devices* **59**, 791–799 (2012).

36. Strelkov, N. *et al.* Impact of Joule heating on the stability phase diagrams of perpendicular magnetic tunnel junctions. *Physical Review B* **98**, 214410 (2018).

37. Sengupta, A., Shim, Y. & Roy, K. Proposal for an all-spin artificial neural network: Emulating neural and synaptic functionalities through domain wall motion in ferromagnets. *IEEE transactions on biomedical circuits and systems* **10**, 1152–1160 (2016).

38. Borders, W. A. *et al.* Integer factorization using stochastic magnetic tunnel junctions. *Nature* **573**, 390–393 (2019).

39. Thermally assisted magnetization reversal in the presence of a spin-transfer torque. *arXiv preprint cond-mat/0302339* (2003).

40. Bedau, D. *et al.* Spin-transfer pulse switching: From the dynamic to the thermally activated regime. *Applied Physics Letters* **97**, (2010).

41. Owen, A. M., McMillan, K. M., Laird, A. R. & Bullmore, E. N-back working memory paradigm: A meta-analysis of normative functional neuroimaging studies. *Human brain mapping* **25**, 46–59 (2005).

42. Jaeggi, S. M., Buschkuehl, M., Perrig, W. J. & Meier, B. The concurrent validity of the

N-back task as a working memory measure. *Memory* **18**, 394–412 (2010).

43. Lamichhane, B., Westbrook, A., Cole, M. W. & Braver, T. S. Exploring brain-behavior relationships in the N-back task. *NeuroImage* **212**, 116683 (2020).
44. Sakai, K., Rowe, J. B. & Passingham, R. E. Active maintenance in prefrontal area 46 creates distractor-resistant memory. *Nature neuroscience* **5**, 479–484 (2002).
45. Jonides, J. & Nee, D. E. Brain mechanisms of proactive interference in working memory. *Neuroscience* **139**, 181–193 (2006).
46. Dewar, M. T., Cowan, N. & Della Sala, S. Forgetting due to retroactive interference: A fusion of Müller and Pilzecker's (1900) early insights into everyday forgetting and recent research on anterograde amnesia. *Cortex* **43**, 616–634 (2007).
47. Tulving, E. & Psotka, J. Retroactive inhibition in free recall: Inaccessibility of information available in the memory store. *Journal of experimental Psychology* **87**, 1 (1971).
48. Glanzer, M. & Cunitz, A. R. Two storage mechanisms in free recall. *Journal of verbal learning and verbal behavior* **5**, 351–360 (1966).
49. Luck, S. J. & Vogel, E. K. The capacity of visual working memory for features and conjunctions. *Nature* **390**, 279–281 (1997).
50. Schurgin, M. W. Visual memory, the long and the short of it: A review of visual working memory and long-term memory. *Attention, Perception, & Psychophysics* **80**, 1035–1056 (2018).
51. He, K., Zhang, X., Ren, S. & Sun, J. Deep residual learning for image recognition. in *Proceedings of the IEEE conference on computer vision and pattern recognition* 770–778

(2016).

52. Fang, W. *et al.* Deep residual learning in spiking neural networks. *Advances in neural information processing systems* **34**, 21056–21069 (2021).

53. Amir, A. *et al.* A low power, fully event-based gesture recognition system. in *Proceedings of the IEEE conference on computer vision and pattern recognition* 7243–7252 (2017).

54. Woo, S. *et al.* Convnext v2: Co-designing and scaling convnets with masked autoencoders. in *Proceedings of the IEEE/CVF conference on computer vision and pattern recognition* 16133–16142 (2023).

55. Wang, Q. *et al.* Dailydvs-200: A comprehensive benchmark dataset for event-based action recognition. in *European Conference on Computer Vision* 55–72 (Springer, 2024).

56. Wang, X. *et al.* Hardvs: Revisiting human activity recognition with dynamic vision sensors. in *Proceedings of the AAAI conference on artificial intelligence* vol. 38 5615–5623 (2024).

57. Hu, Y., Binas, J., Neil, D., Liu, S.-C. & Delbruck, T. Ddd20 end-to-end event camera driving dataset: Fusing frames and events with deep learning for improved steering prediction. in *2020 IEEE 23rd International Conference on Intelligent Transportation Systems (ITSC)* 1–6 (IEEE, 2020).

58. Madjid, N. A. *et al.* Trajectory prediction for autonomous driving: Progress, limitations, and future directions. *arXiv preprint arXiv:2503.03262* (2025).

59. Maqueda, A. I., Loquercio, A., Gallego, G., García, N. & Scaramuzza, D. Event-based vision meets deep learning on steering prediction for self-driving cars. in *Proceedings of the IEEE conference on computer vision and pattern recognition* 5419–5427 (2018).

60. De Haan, P., Jayaraman, D. & Levine, S. Causal confusion in imitation learning. *Advances in neural information processing systems* **32**, (2019).

61. Wen, C., Lin, J., Darrell, T., Jayaraman, D. & Gao, Y. Fighting copycat agents in behavioral cloning from observation histories. *Advances in Neural Information Processing Systems* **33**, 2564–2575 (2020).

62. Willmore, B. D. & King, A. J. Adaptation in auditory processing. *Physiological Reviews* **103**, 1025–1058 (2023).

63. Kohn, A. Visual adaptation: physiology, mechanisms, and functional benefits. *Journal of neurophysiology* **97**, 3155–3164 (2007).

64. Moser, T., Grabner, C. P. & Schmitz, F. Sensory processing at ribbon synapses in the retina and the cochlea. *Physiological reviews* **100**, 103–144 (2020).

65. Tuma, T., Pantazi, A., Le Gallo, M., Sebastian, A. & Eleftheriou, E. Stochastic phase-change neurons. *Nature nanotechnology* **11**, 693–699 (2016).

66. Wang, Z. *et al.* Memristors with diffusive dynamics as synaptic emulators for neuromorphic computing. *Nature materials* **16**, 101–108 (2017).

67. Mulaosmanovic, H., Mikolajick, T. & Slesazeck, S. Accumulative polarization reversal in nanoscale ferroelectric transistors. *ACS applied materials & interfaces* **10**, 23997–24002 (2018).

68. Liu, H. & Ohsawa, T. A binarized spiking neural network based on auto-reset LIF neurons and large signal synapses using STT-MTJs. *Japanese journal of applied physics* **62**, 044501 (2023).

69. Yan, B., Yang, Y. & Huang, R. Memristive dynamics enabled neuromorphic computing

systems. *Sci. China Inf. Sci.* **66**, 200401 (2023).

70. Tatsumura, K., Yazdanshenas, S. & Betz, V. High density, low energy, magnetic tunnel junction based block RAMs for memory-rich FPGAs. in *2016 International Conference on Field-Programmable Technology (FPT)* 4–11 (2016). doi:10.1109/FPT.2016.7929181.
71. Zhao, W. *et al.* Cross-Point Architecture for Spin-Transfer Torque Magnetic Random Access Memory. *IEEE Transactions on Nanotechnology* **11**, 907–917 (2012).
72. Fang, W. *et al.* Spikingjelly: An open-source machine learning infrastructure platform for spike-based intelligence. *Science Advances* **9**, eadi1480 (2023).
73. Liu, Z. *et al.* Swin transformer: Hierarchical vision transformer using shifted windows. in *Proceedings of the IEEE/CVF international conference on computer vision* 10012–10022 (2021).

## Acknowledgements

We thank the Novel IC Exploration Facility (NICE) and Haoxun Luo for their assistance with electrical characterization. We also acknowledge the Suzhou Institute of Nano-Tech and Nano-Bionics, CAS, for their support in device fabrication.

## Funding

Guangzhou-HKUST(GZ) Joint Funding Program (No. 2025A03J3717)

## Author contributions

J.L. and K.Y. conceived the core concept and designed the study. K.Y. supervised the project. J.L. performed the electrical characterization, developed the algorithms, and collected the data. H.Z. assisted with algorithm implementation and contributed to scientific discussions. B.Z. assisted with the validation of the algorithms. J.L. wrote the manuscript, and K.Y. provided critical revisions.

## Competing interests

The authors declare no competing interests.

## Data availability

The datasets generated during and analysed during the current study are available from the corresponding author on reasonable request.

## Code availability

The implementation code will be made publicly available on GitHub upon acceptance of the manuscript. During the review process, the code is available from the corresponding author upon reasonable request.

## Supplementary Information

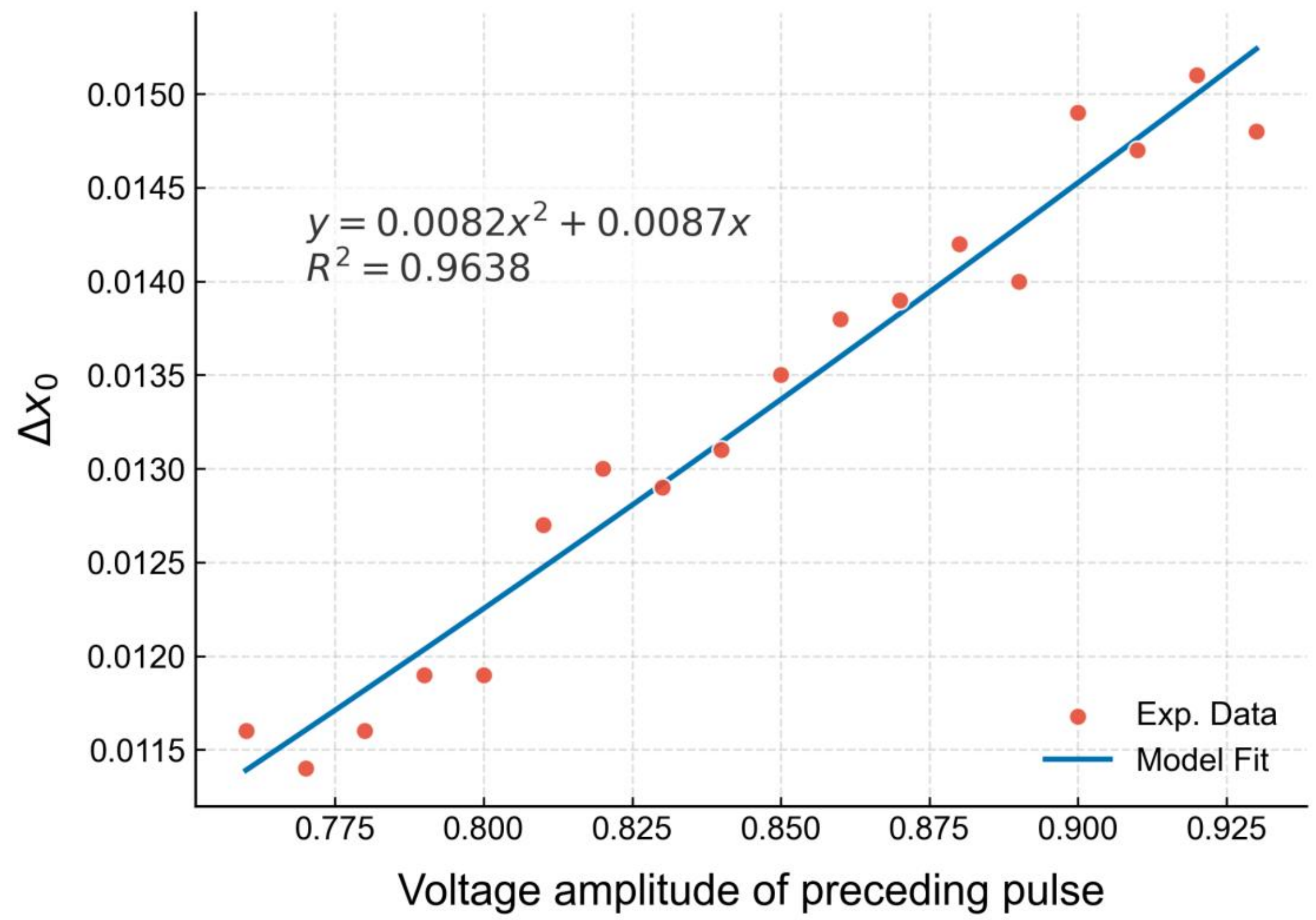

**Supplementary Fig. 1. Dependence of the sigmoid mean parameter shift $\Delta x_0$ on the preceding pulse amplitude.** The scatter plot shows the experimentally measured shift in the sigmoid mean parameter $\Delta x_0$ as a function of the voltage amplitude of the preceding pulse (orange solid dots). The relationship is characterized by a quadratic increase, as indicated by the fitted curve (blue line; $y = 0.0082x^2 + 0.0087x$). The model yields a coefficient of determination $R^2 = 0.9638$, demonstrating a high goodness of fit.

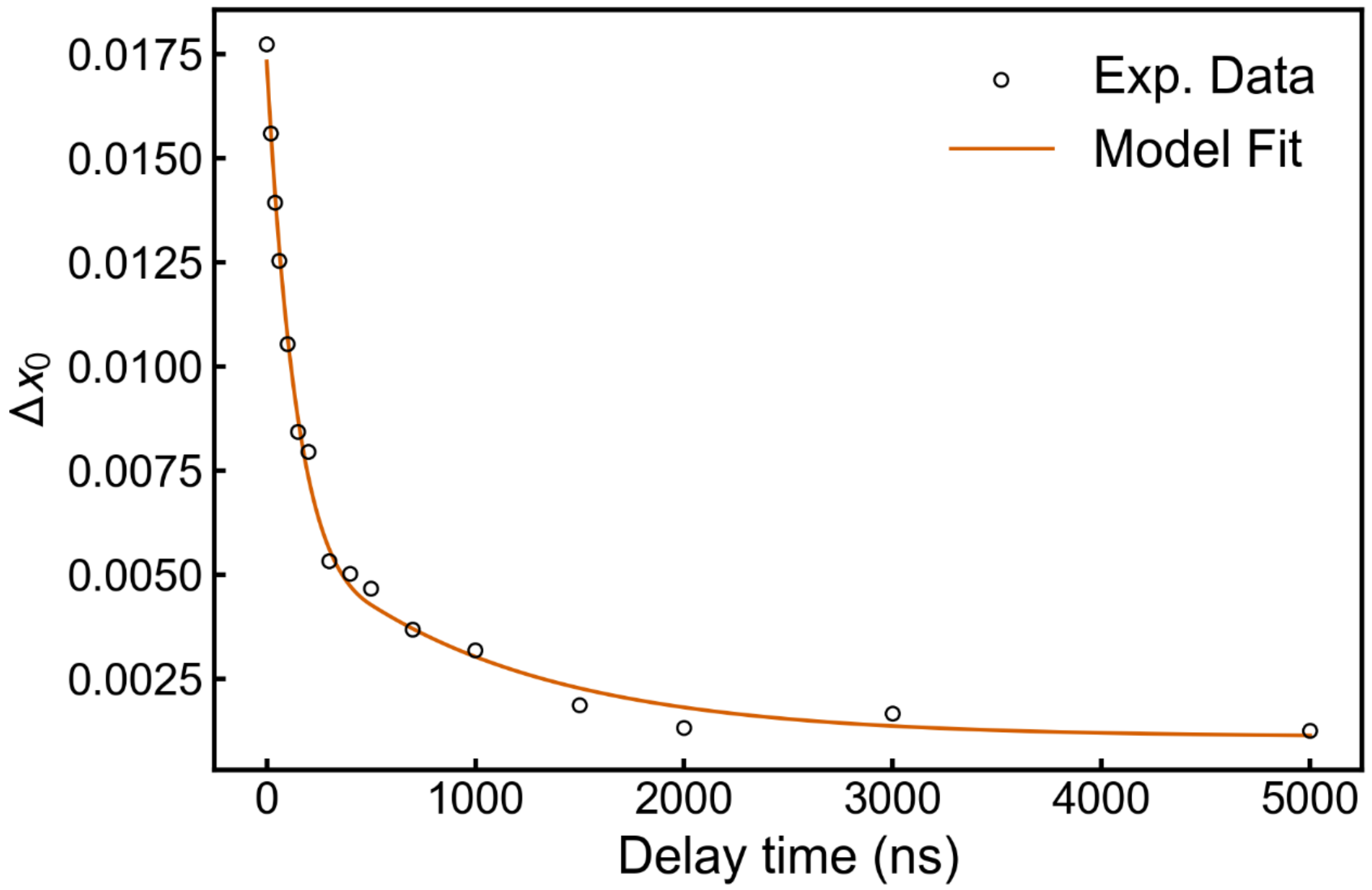

**Supplementary Fig. 2. Temporal decay of the sigmoid mean parameter shift $\Delta x_0$.** The data show the relaxation of $x_0$ following a train of five voltage pulses (0.9 V amplitude). The solid orange line represents the best fit using a two-segment piecewise exponential model with a breakpoint at 500 ns.

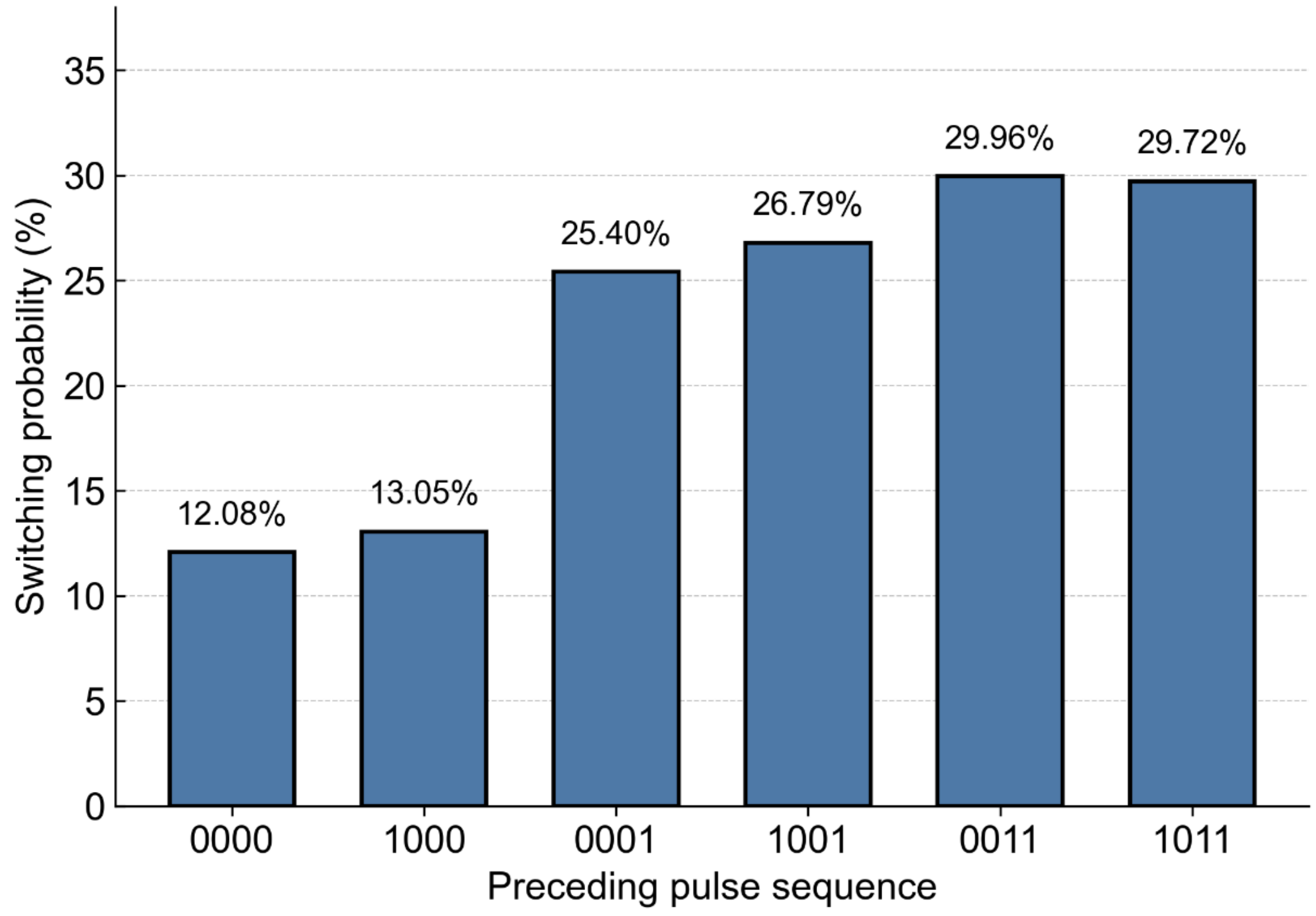

**Supplementary Fig. 3. Dependence of MTJ switching probability on the preceding pulse sequence.** The input consists of a four-step sequence where '1' and '0' denote the presence and absence of an input spike (0.85 V, 230 ns period), respectively. Switching probability was assessed using a probe pulse (0.83 V, 150 ns) applied immediately after the sequence. The digits follow chronological order. For instance, '1000' indicates a pulse at the first time step followed by three idle steps.

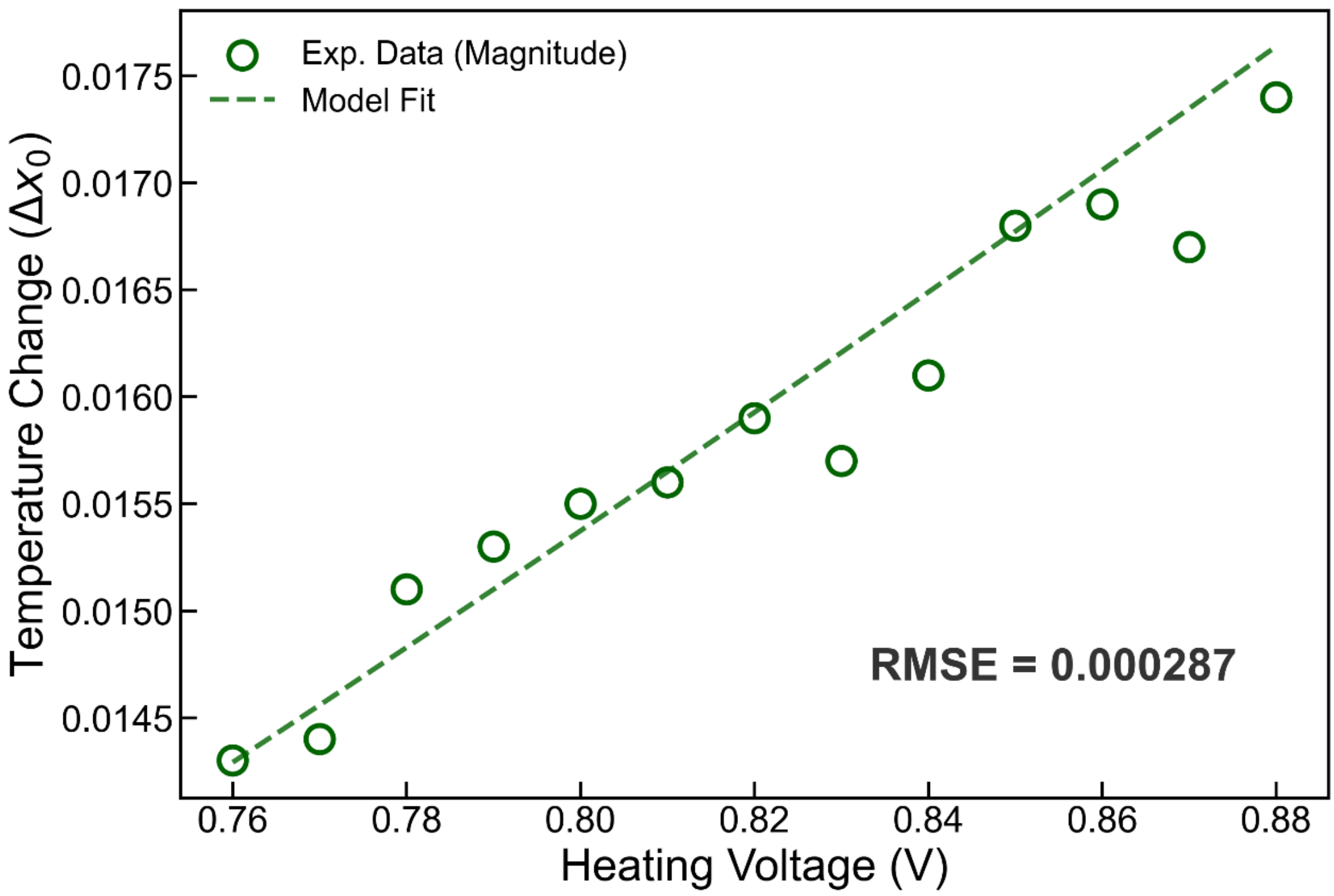

**Supplementary Fig. 4. Validation of the thermal intrinsic plasticity algorithmic model.** The plot compares the modeled (dashed line) and experimentally measured (open circles) temperature change ($\Delta x_0$) of the MTJ following three pre-pulses (230 ns period) at varying voltages. The $\Delta x_0$ values were derived from the switching probability measured using a probe pulse (0.83 V, 150 ns) (see Methods). The algorithmic simulation agrees well with the experimental data, with a root-mean-square error (RMSE) of 0.000287.

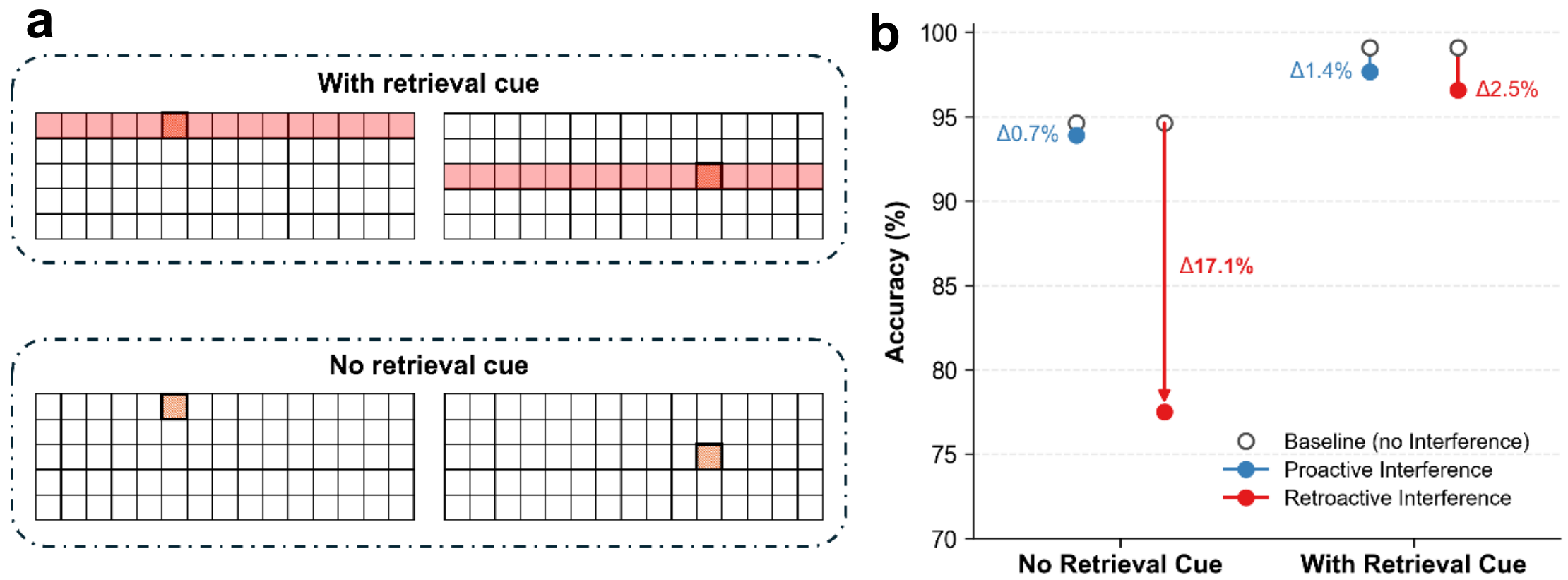

**Supplementary Fig. 5. a**, Schematic representation of the cued versus uncued retrieval paradigms. The memory space consists of 50 classes arranged in a 5×10 grid (analogous to the low-similarity condition in Fig. 3). In the cued condition (top), the specific row containing the target is provided as a contextual input (highlighted in pink), whereas no spatial hint is available in the uncued condition (bottom). **b**, Quantitative comparison of classification accuracy under Proactive Interference (PI) and Retroactive Interference (RI). While PI (blue solid circles) exerts negligible impact in both scenarios, the significant performance deficit induced by RI (red solid circles, Δ17.1%) in the uncued condition is largely ameliorated by the presence of the retrieval cue, reducing the deficit to Δ2.5%. White open circles denote the interference-free baseline. The temporal sequence and experimental protocols are identical to the interference experiments in Fig. 3 (see Methods).

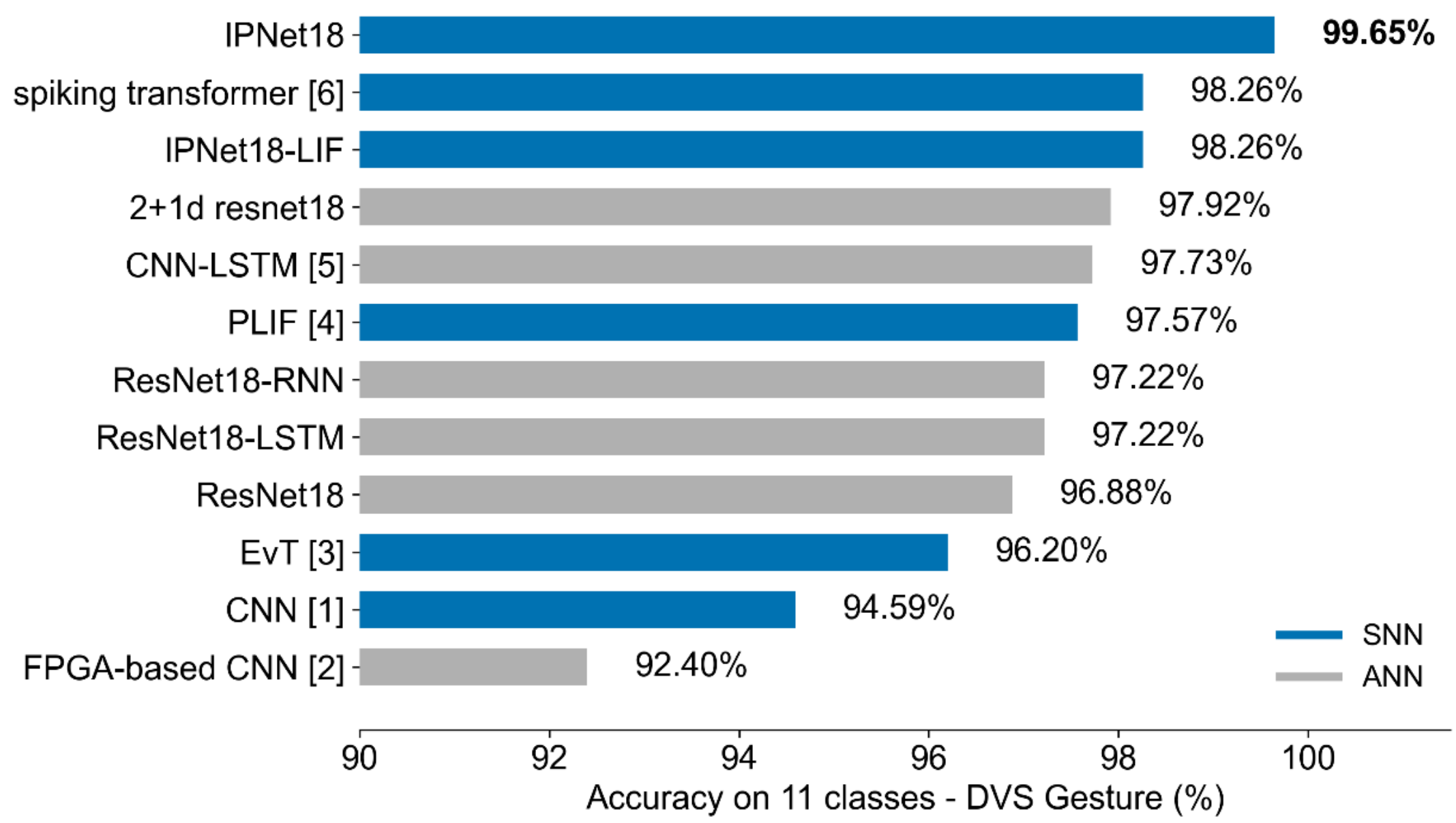

**Supplementary Fig. 6. Performance benchmarking on the 11-class DVS Gesture dataset.** The bar chart compares the classification accuracy of IPNet18 with other reported models[1–6] on the 11-class DVS Gesture task. The IPNet18 achieves a leading accuracy of 99.65%. Blue and gray bars denote models based on Spiking Neural Networks (SNNs) and Artificial Neural Networks (ANNs), respectively.

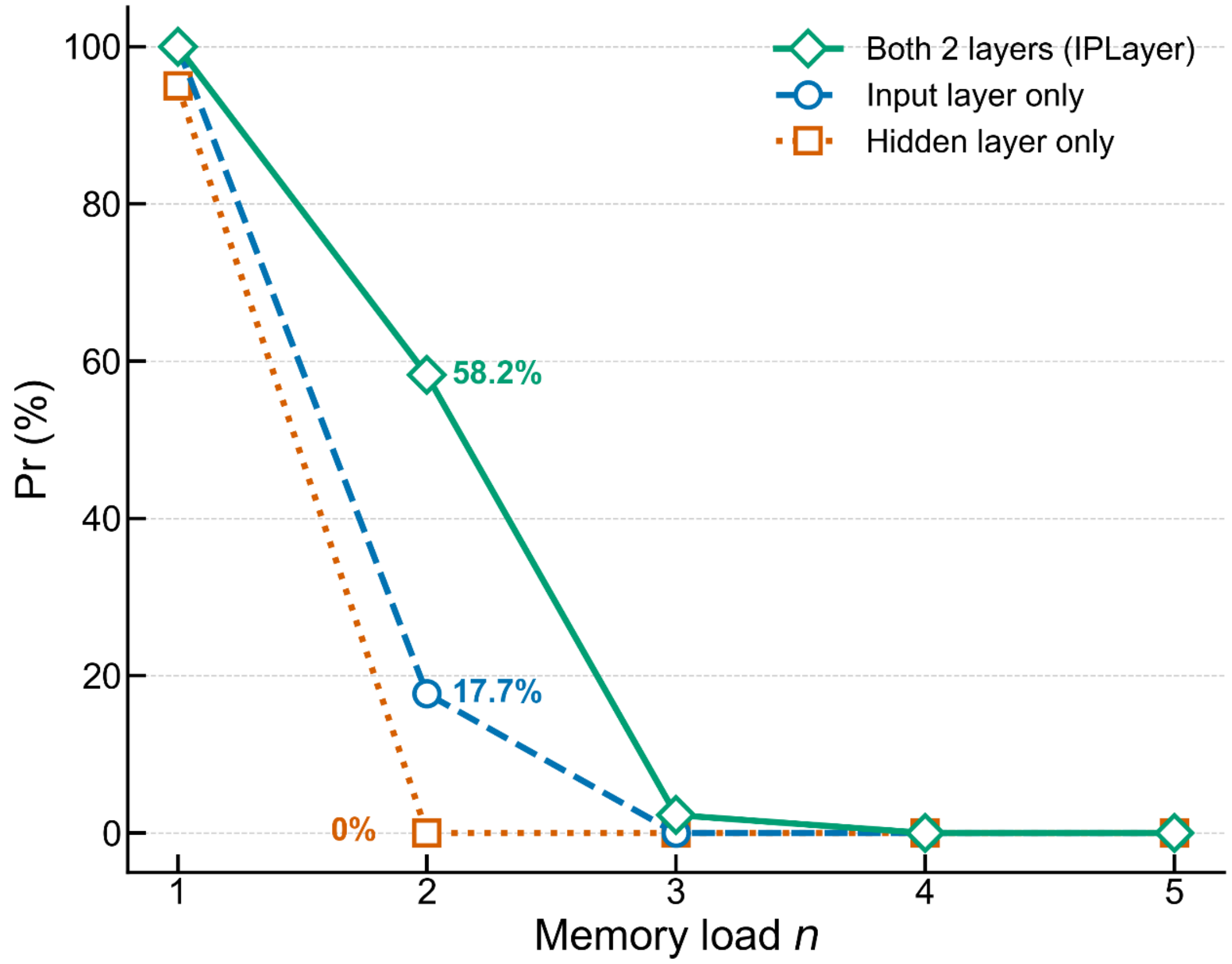

**Supplementary Fig. 7. Performance of the FCN-based IPNet on the n-back task with varying IPLayer placements.** The model is built on a fully connected network (FCN) topology with 225 input, 225 hidden, and 2 output neurons. Performance rate ($P_r$ (%)) is plotted as a function of memory load $n$. The green solid line with diamonds denotes the configuration where standard neurons in both the input and hidden layers are replaced by IPNeurons ("Both 2 layers"). The blue dashed line with circles and the orange dotted line with squares represent models with IPNeurons implemented only in the input layer ("Input layer only") or the hidden layer ("Hidden layer only "), respectively. Annotated percentages indicate the $P_r$ at $n$=2.

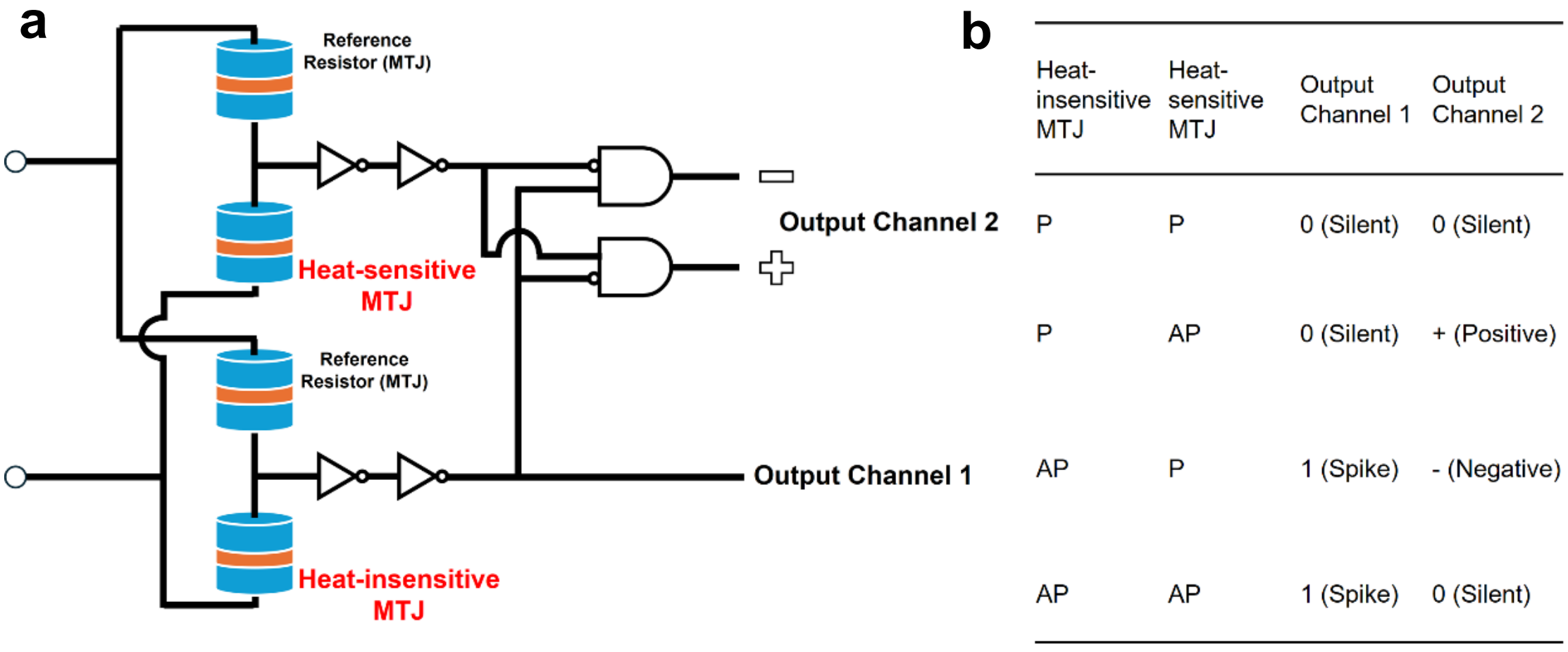

| Heat-insensitive MTJ | Heat-sensitive MTJ | Output Channel 1 | Output Channel 2 |
| --- | --- | --- | --- |
| P | P | 0 (Silent) | 0 (Silent) |
| P | AP | 0 (Silent) | + (Positive) |
| AP | P | 1 (Spike) | - (Negative) |
| AP | AP | 1 (Spike) | 0 (Silent) |

**Supplementary Fig. 8. Schematic of the MTJ-based differential intrinsic plasticity neuron circuit. a**, Schematic of the circuit comprising two active MTJs (text in red), two reference MTJs, and peripheral CMOS. The active MTJs differ in thermal sensitivity (heat-sensitive vs. heat-insensitive). **b**, Truth table summarizing the output logic. Output Channel 1 depends solely on the heat-insensitive MTJ state (firing in the antiparallel/AP state; silent in the parallel/P state). Output Channel 2 generates ternary outputs (Positive/Negative/Zero) based on the combined resistance states of both active MTJs. In this digital implementation, two physical output lines are used to represent positive and negative spikes, respectively.

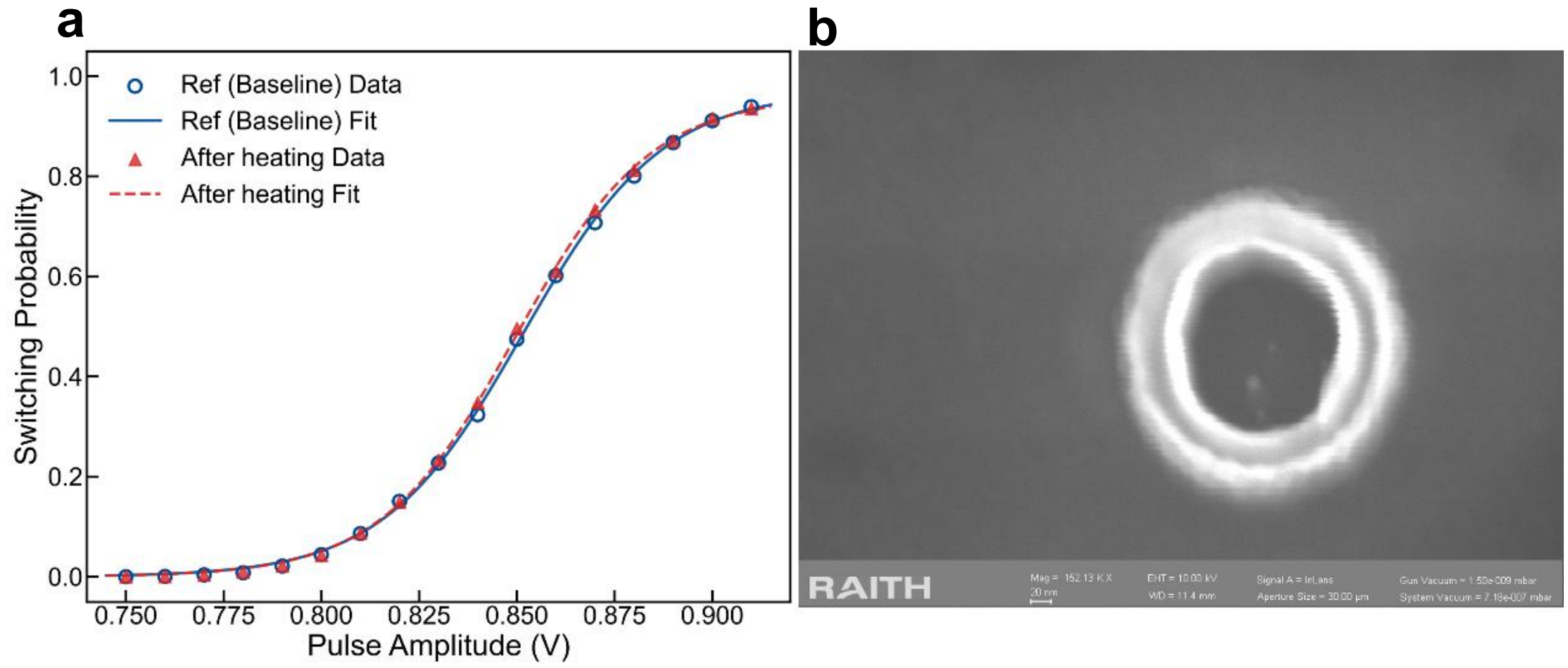

**Supplementary Fig. 9. Characterization of thermal sensitivity in a 100 nm MTJ.** **a**, Switching probability as a function of pulse amplitude. Blue circles represent baseline data collected without pre-pulses, while red triangles denote measurements taken immediately after three pulses (0.9 V SET, 230 ns total cycle). The solid and dashed lines indicate the corresponding sigmoid fits. The overlapping curves demonstrate that the switching probability of the 100 nm MTJ is insensitive to the heating protocol, in sharp contrast to the behavior of the 200 nm MTJ shown in Fig. 2f, g. **b**, Top-view scanning electron microscopy (SEM) image of a 100 nm MTJ.

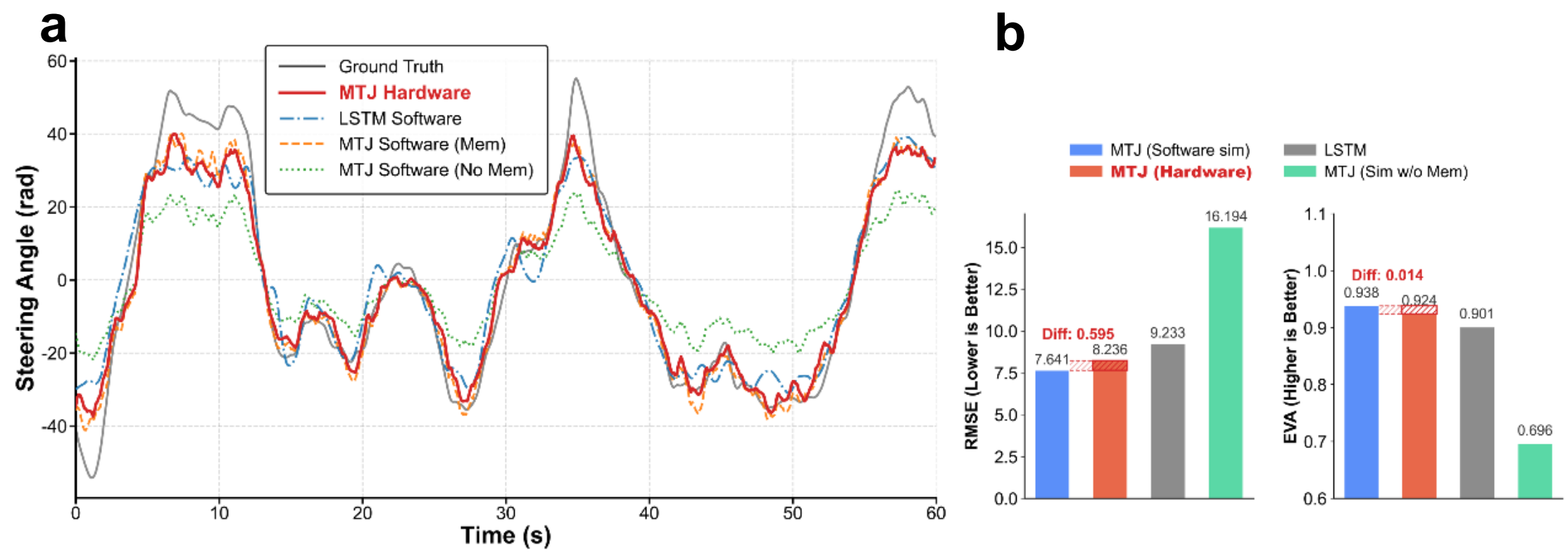

**Supplementary Fig. 10. Additional hardware-in-the-loop evaluation on a distinct driving segment. a**, Steering angle trajectories over an additional 600-frame test sequence (separate from the segment in Fig. 6). **b**, Quantitative metrics (RMSE and EVA) for this 600 frames segment.

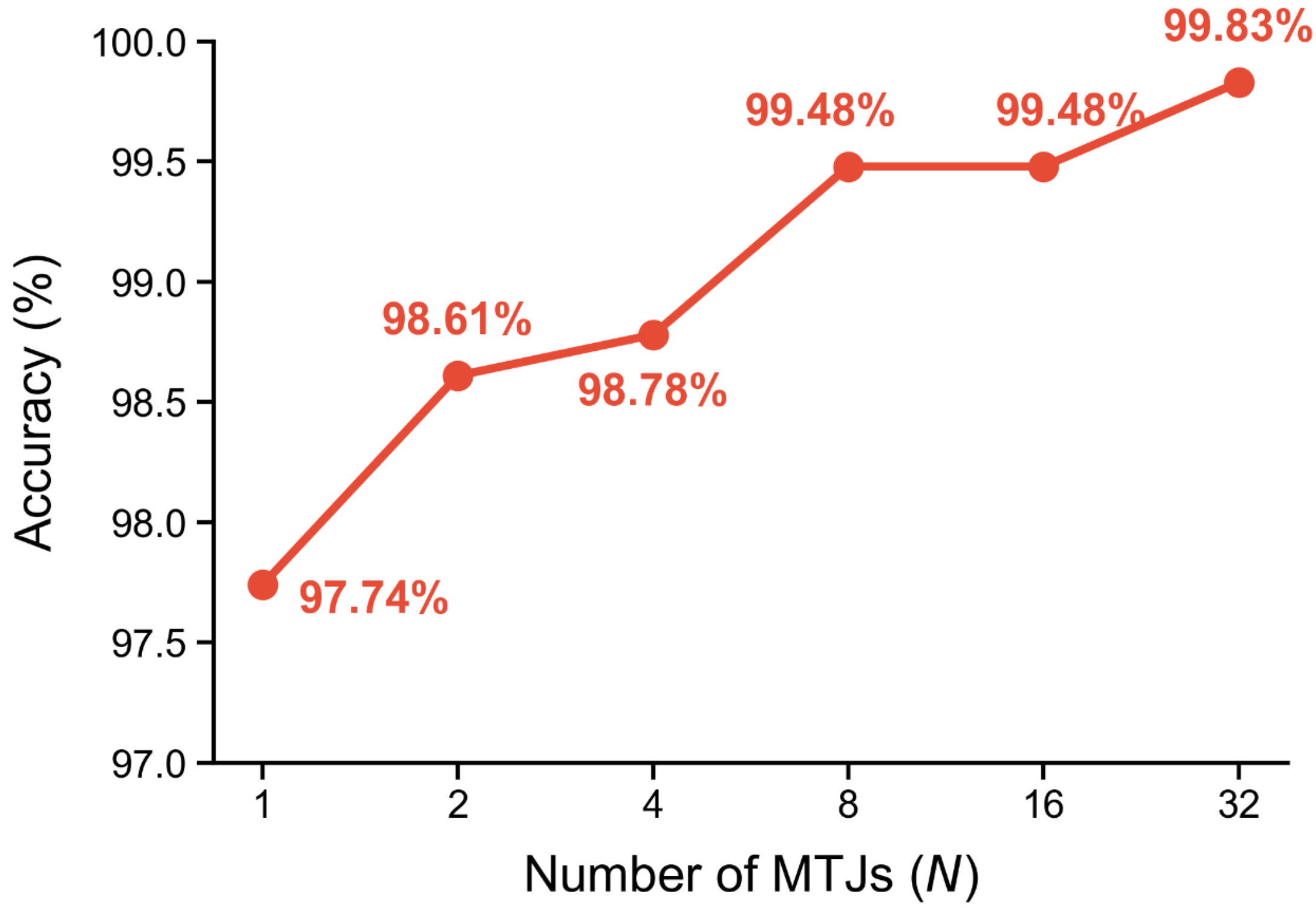

**Supplementary Fig. 11. Dependence of classification accuracy on the ensemble size of thermally sensitive MTJs (*N*) per computing unit.** The parameter $N$ denotes the number of thermally sensitive MTJs integrated within a single computing unit. This configuration allows the unit to perform average pooling across $N$ artificial neurons, directly mapping the hardware parallelism to $N$ Monte Carlo sampling iterations per time step. In this experiment on the Time-Reversed DVS Gesture task, the network was trained using only the initial 20% of the video frames but tested on the full sequence. Notably, even a minimal configuration ($N = 1$) yields an accuracy of 97.74%, surpassing conventional LSTM and 3D-CNN baselines. Increasing the sampling density improves performance, reaching the highest accuracy of 99.83% at $N = 32$.

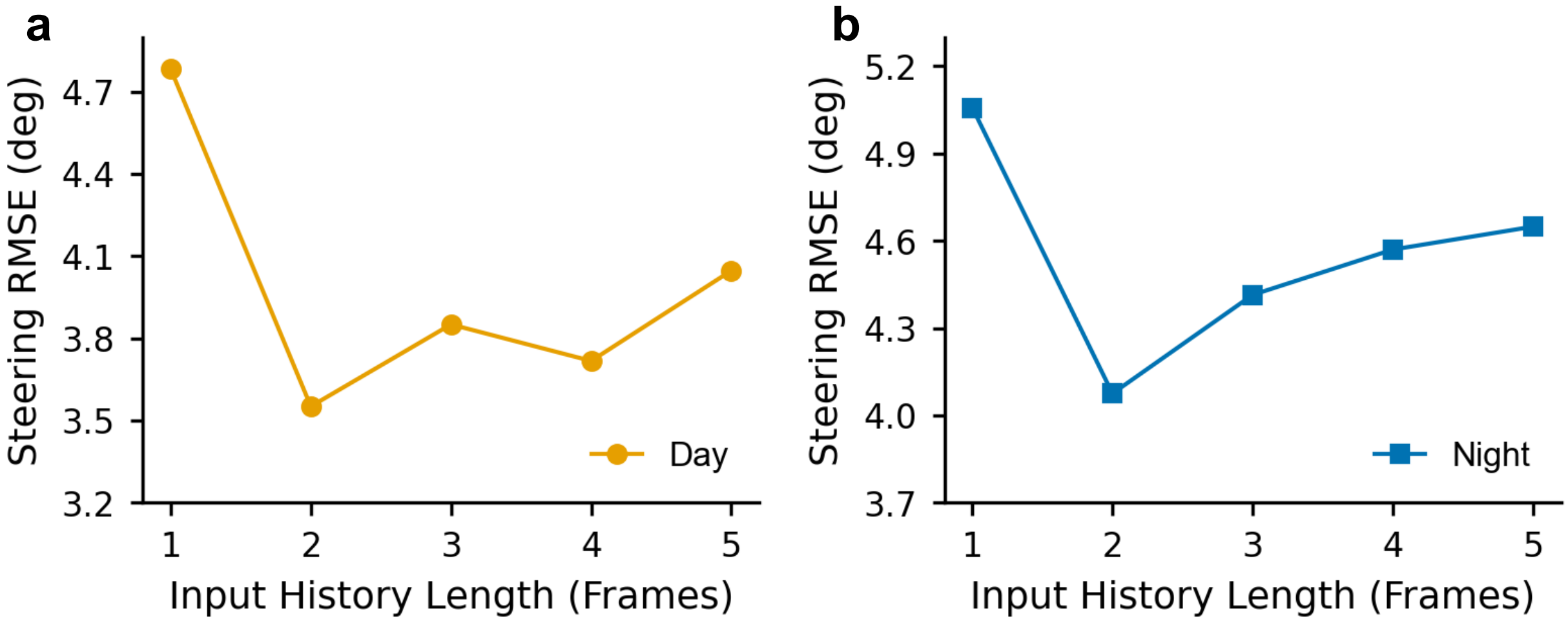

**Supplementary Fig. 12. Steering performance of the Video Swin Transformer across varying input history lengths.** Steering Root Mean Square Error (RMSE) is plotted as a function of the input history length (in frames) under Day **(a)** and Night **(b)** illumination conditions. Consistent with the LSTM models shown in the main text, this state-of-the-art attention-based architecture exhibits a U-shaped error profile.

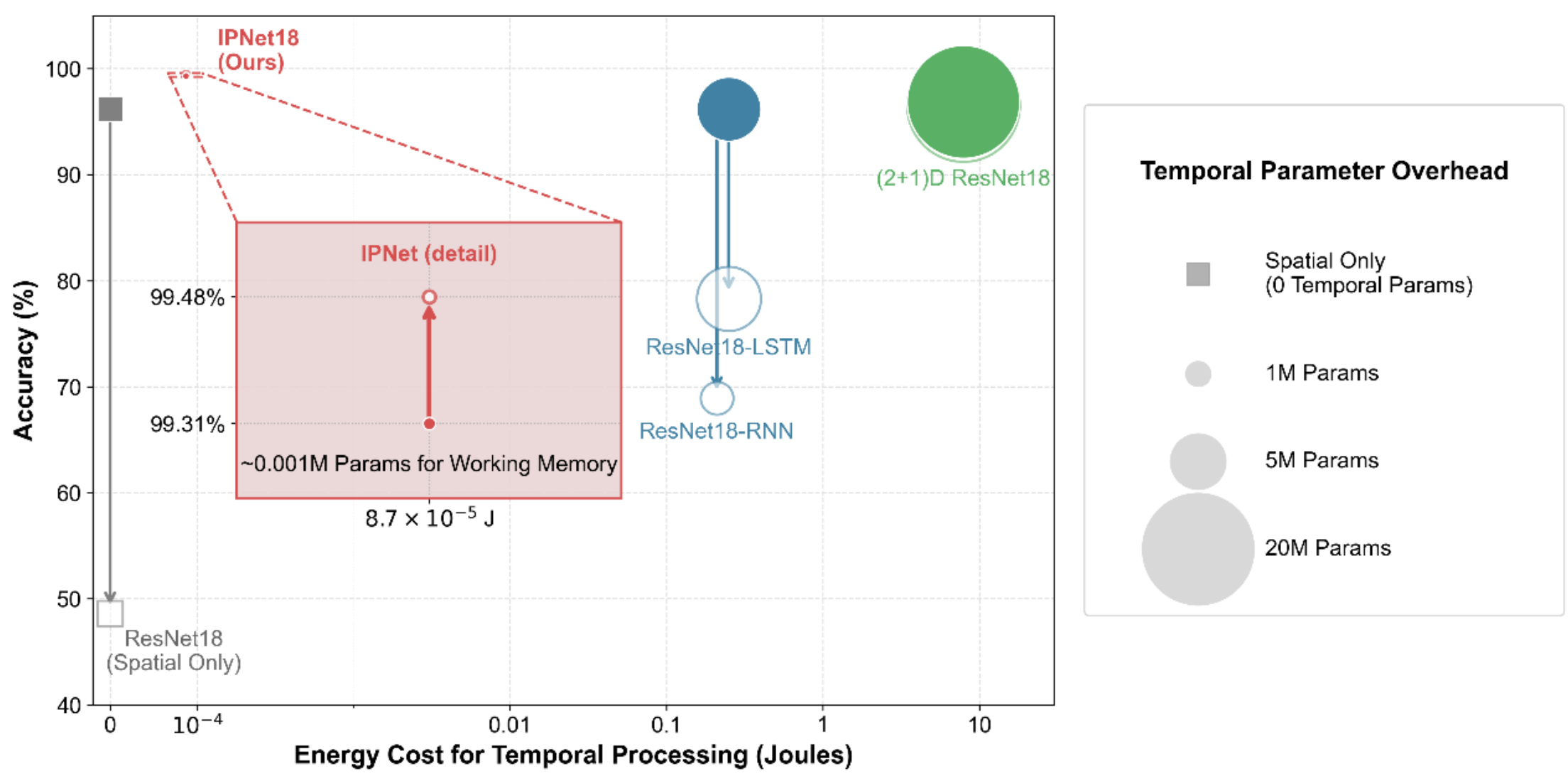

**Supplementary Fig. 13. Accuracy-efficiency landscape for temporal processing under restricted input history.** Accuracy-efficiency comparison of IPNet18, ResNet18-RNN, ResNet18-LSTM and R(2+1)D-ResNet18, with a spatial-only ResNet18 as the reference. Filled markers indicate performance when the complete input sequence is available, whereas open markers indicate performance when only the first 20% of frames are available. Squares denote the spatial-only baseline and circles denote models incorporating temporal processing. Marker area is proportional to the number of additional parameters introduced for temporal processing relative to the ResNet18 backbone. The shaded callout highlights the low-overhead operating regime of IPNet18: its working-memory mechanism requires approximately 0.001 million additional parameters and an incremental energy cost of $8.7\times10^{-5}$J per sample, while changing accuracy from 99.31% to 99.48%. Energy and parameter overheads for temporal processing are calculated relative to the spatial-only ResNet18 baseline.

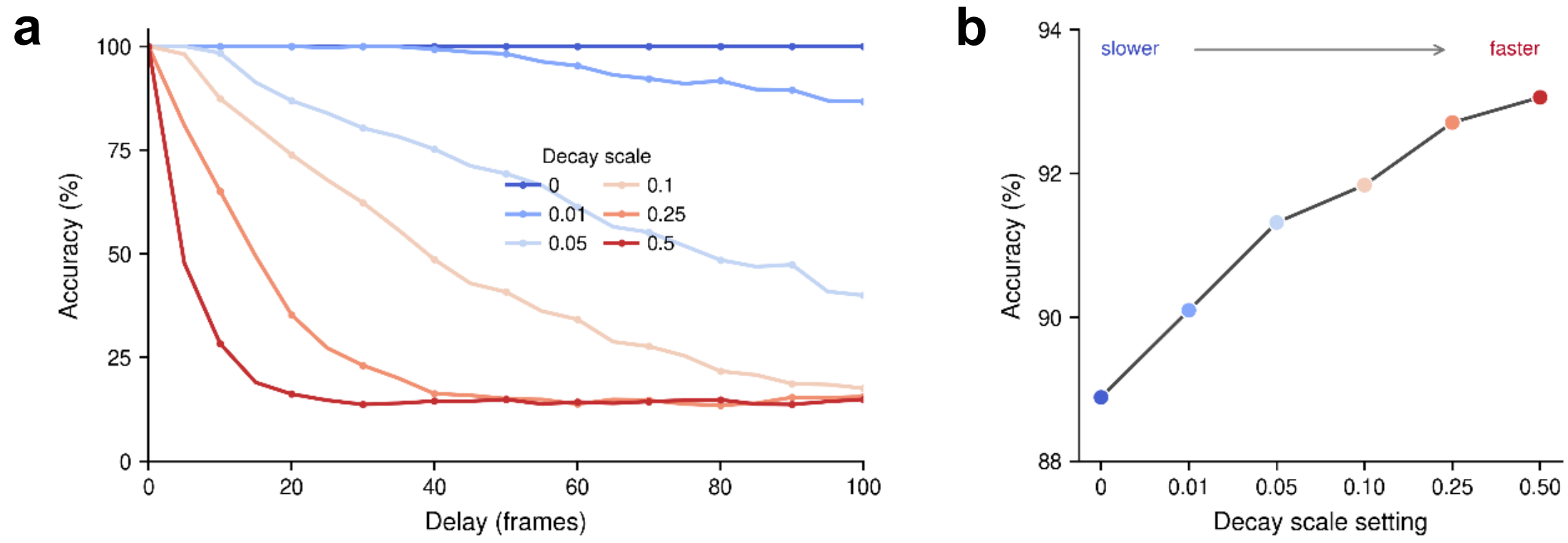

**Supplementary Fig. 14. Task-dependent effects of intrinsic-state decay. a,** Delay-matching accuracy as a function of delay for simulated MTJ neurons with different decay scales. Smaller decay scales preserved performance over longer delays; a decay scale of 0 represents no decay. **b,** Accuracy on the Time-Reversed DVS Gesture task at six decay scale settings, shown at equal spacing. Models were trained using the first 20% of each sequence and evaluated on the complete sequence. Accuracy increased monotonically with decay scale.

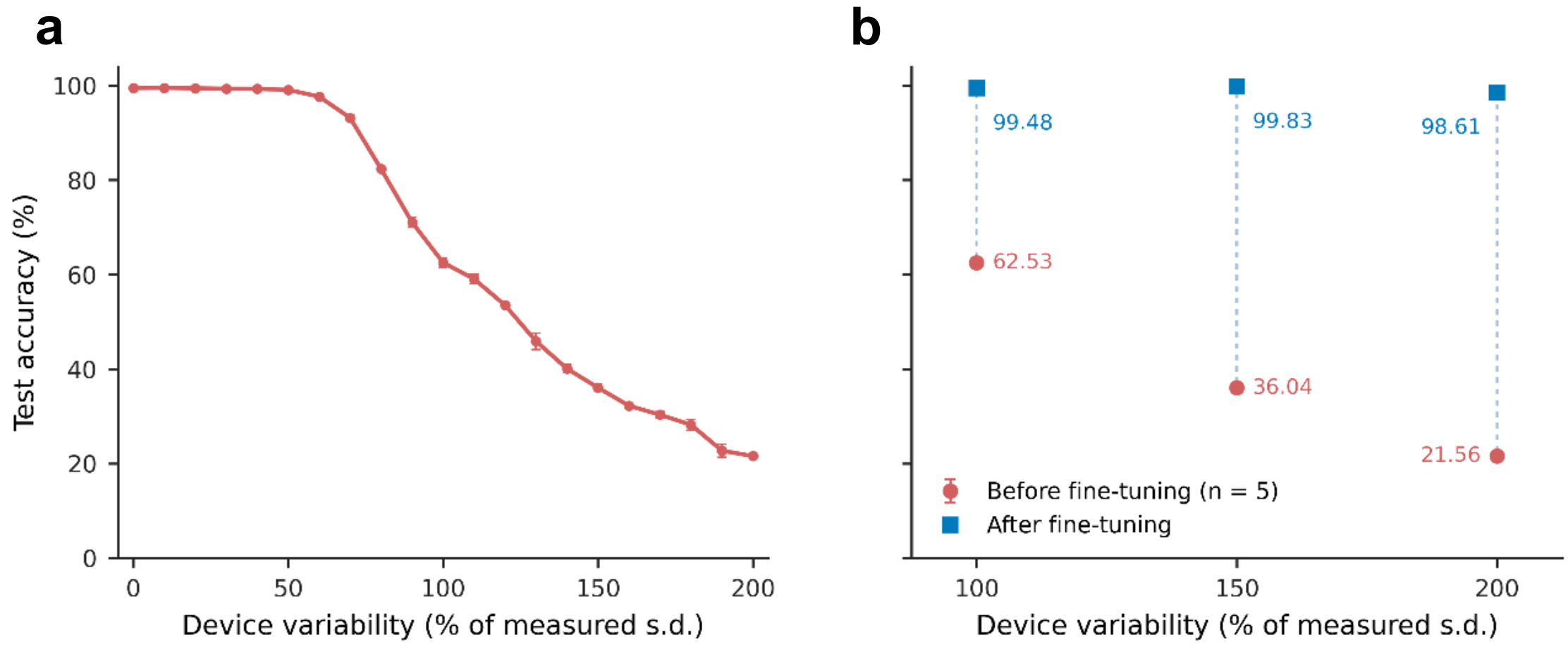

**Supplementary Fig. 15. Robustness of IPNet to device-to-device variability and recovery through device-aware fine-tuning. a**, Accuracy on the Time-Reversed DVS Gesture task with increasing device-parameter variability. Data are mean ± s.d. from five randomly sampled device arrays. **b**, Accuracy before and after device-aware fine-tuning at variability levels of 100%, 150% and 200%. Dashed lines connect results at the same variability level; fine-tuned results are from single runs.

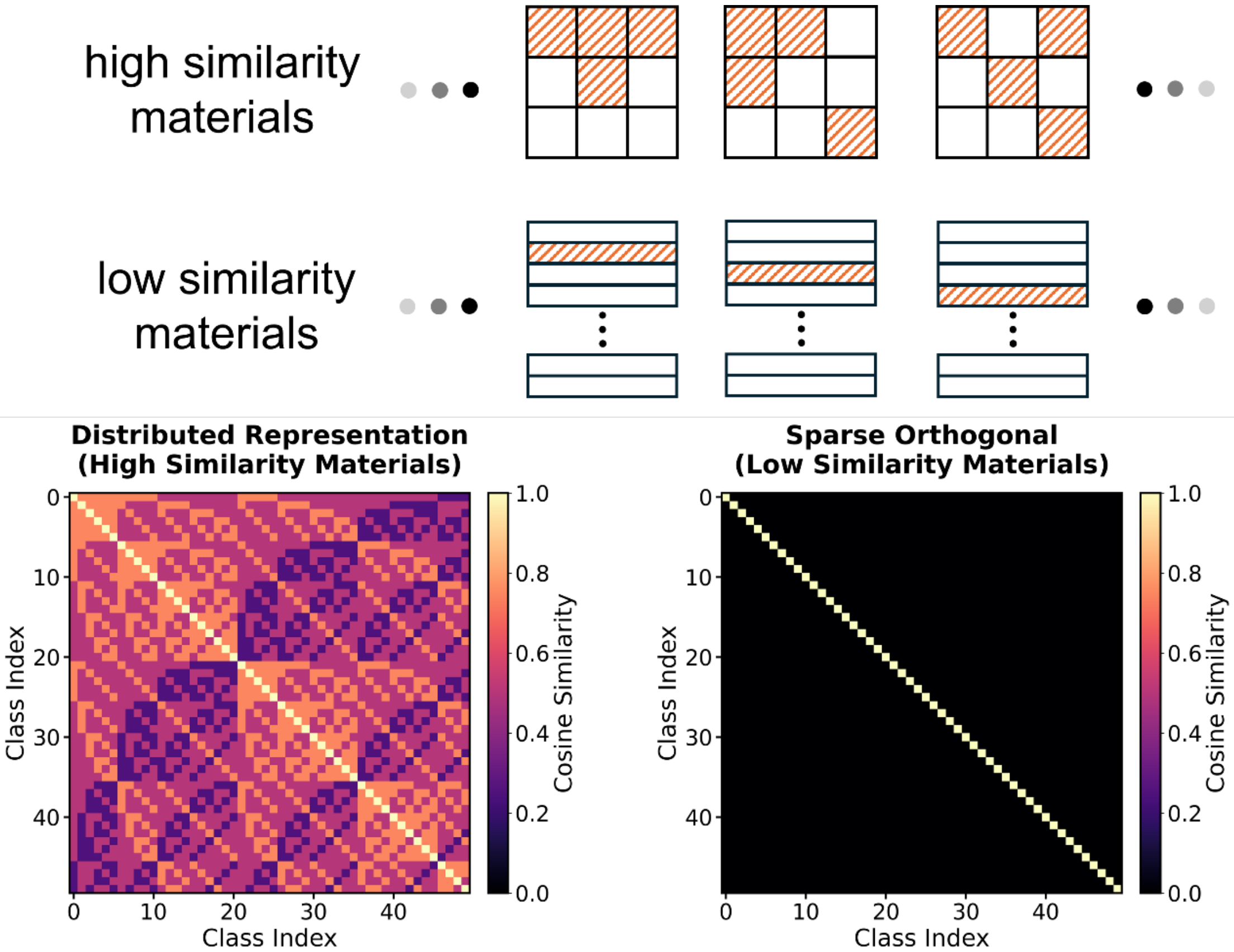

**Supplementary Fig. 16. Comparison of material similarity representations.** The upper panels visualize the encoding strategies: high-similarity materials share overlapping grid units (distributed representation), while low-similarity the corresponding cosine similarity matrices. The high-similarity materials exhibit significant off-diagonal correlations, whereas the low-similarity materials show zero cross-item similarity (indicating orthogonality).

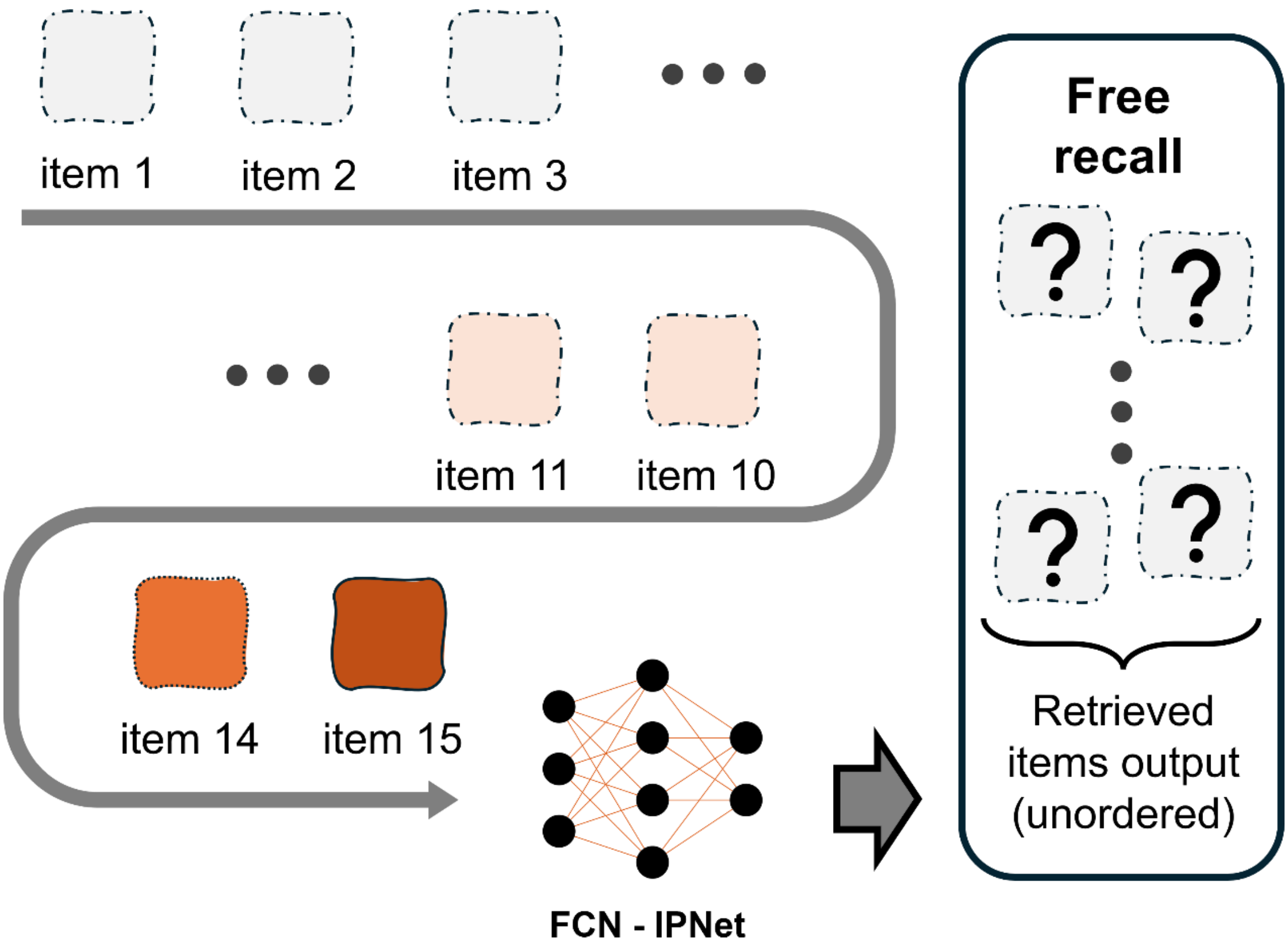

**Supplementary Fig. 17. Schematic of the free recall task.** A sequence of 15 distinct items, randomly sampled from a pool of 50 stimuli, is presented serially to the FCN-IPNet. After the presentation phase, the model reports the set of presented items without an order constraint.

| Device parameter | Symbol/code parameter | Nominal value | Characterized s.d. | Relative s.d. |
|---|---|---|---|---|
| Maximum switching probability | $(L)$ | 0.9186 | 0.067231 | 7.32% |
| Sigmoid slope | $(k)$ | 83.2776 | 24.918801 | 29.92% |
| Ambient sigmoid midpoint | $(x_{env})$ | 0.8541 | 0.029317 | 3.43% |
| Voltage-dependent heating coefficient | $(b_{heating})$ | 0.0087 | 0.004095 | 47.07% |
| Cooling-rate coefficient | $(r_{cooling})$ | 0.0067096 | 0.000992 | 14.78% |

**Supplementary Table 1. Experimentally characterized device-parameter variability used in the device-to-device variability simulations.** The nominal parameter values were extracted from a representative MTJ and used to configure the IP-neuron model during network training. The standard deviations were derived from the experimentally characterized device-to-device distributions of the 200-nm-diameter MTJs and define a variability scale of 100%. Relative standard deviations were calculated as the characterized standard deviation divided by the absolute nominal value. During inference, the parameters of individual IP neurons were sampled independently from these distributions. Sampled parameters were constrained to their physically valid ranges.

## Supplementary References

1. Fu, Z. & Ye, W. A 593nj/inference dvs hand gesture recognition processor embedded with reconfigurable multiple constant multiplication technique. *IEEE Transactions on Circuits and Systems I: Regular Papers* **71**, 2749–2759 (2024).
2. Amir, A. *et al.* A low power, fully event-based gesture recognition system. in *Proceedings of the IEEE conference on computer vision and pattern recognition* 7243–7252 (2017).
3. Sabater, A., Montesano, L. & Murillo, A. C. Event transformer. a sparse-aware solution for efficient event data processing. in *Proceedings of the IEEE/CVF Conference on Computer Vision and Pattern Recognition* 2677–2686 (2022).
4. Fang, W. *et al.* Incorporating learnable membrane time constant to enhance learning of spiking neural networks. in *Proceedings of the IEEE/CVF international conference on computer vision* 2661–2671 (2021).
5. Lin, X., Liu, M. & Chen, H. Spike-HAR++: an energy-efficient and lightweight parallel spiking transformer for event-based human action recognition. *Frontiers in Computational Neuroscience* **18**, 1508297 (2024).
6. Innocenti, S. U., Becattini, F., Pernici, F. & Del Bimbo, A. Temporal binary representation for event-based action recognition. in *2020 25th International Conference on Pattern Recognition (ICPR)* 10426–10432 (IEEE, 2021).